\documentclass[manuscript, screen]{acmart}

\AtBeginDocument{%
  \providecommand\BibTeX{{%
    \normalfont B\kern-0.5em{\scshape i\kern-0.25em b}\kern-0.8em\TeX}}}

\setcopyright{acmlicensed}
\copyrightyear{2024}
\acmYear{2024}
\acmDOI{XXXXXXX.XXXXXXX}

\acmJournal{JACM}
\acmVolume{37}
\acmNumber{4}
\acmArticle{111}
\acmMonth{8}

\acmISBN{978-1-4503-XXXX-X/18/06}

\usepackage{booktabs}
\usepackage{graphicx}
\usepackage{subcaption}
\usepackage[export]{adjustbox}
\usepackage{wrapfig}
\usepackage{pdfpages}
\usepackage{pgfplotstable}
\usepackage{blindtext}
\usepackage{titlesec}

\begin{document}

\title{Exploring the Applications of Generative AI in High School STEM Education}

\author{Ishaan Masilamony}

\renewcommand{\shortauthors}{Masilamony}

\begin{abstract}

    In recent years, ChatGPT \cite{openai_2023_gpt4} along with Microsoft
    Copilot have become subjects of great discourse, particularly in the
    field of education. Prior research has hypothesized on potential
    impacts these tools could have on student learning and performance.
    These have primarily relied on trends from prior applications of
    technology in education and an understanding of the limitations and
    strengths of Generative AI in other applications. This study utilizes
    an experimental approach to analyze the impacts of Generative AI on
    high school STEM education (physics in particular). In accordance with
    most findings, generative AI does have some positive impact on student
    performance. However, our findings have shown that the most
    significant impact is an increase in student engagement with the
    subject.

\end{abstract}

\ccsdesc[500]{Applied computing~Interactive learning environments}
\ccsdesc[300]{Social and professional topics~Computing / technology policy}
\ccsdesc[500]{Applied computing~Computer-assisted instruction}
\ccsdesc[300]{Computing methodologies~Natural language processing}
\ccsdesc[500]{Computing methodologies~Natural language generation}

\keywords{Generative AI, machine learning, education, equity, physics}


\maketitle

\newpage
\tableofcontents
\newpage
\section{Introduction}
Generative AI (GAI) refers to a subset of Artificial Intelligence that
generates novel outputs, such as text, audio, images, and video. This is
not a novel concept; models such as GPT-2 (an early GAI model) have
existed for many years but drew the attention of only the computer science
community due to their limitations in terms of functionality and ease of
use. However, recent advancements in technology and the models themselves,
primarily due to increases in dataset size, have brought GAI and
AI-generated content (AIGC) to the focus of not only the academic
community but also that of greater society \cite{nah_2023_generative}.
ChatGPT has drawn significant attention due to its conversational
capabilities, essentially being able to take a naturally worded prompt and
respond using novel, natural language that is structurally and
syntactically similar to human-written text. Additionally, these newer
models also utilize reinforcement learning from human feedback (RLHF),
which enables them to determine the most appropriate response for a given
prompt based on human feedback, enabling them to improve in accuracy and
reliability over time \cite{cao_2023_a}. Thus, these models have improved
significantly in recent years and have become more relevant than ever to
not only the academic community but to society as a whole.

With the advent of these new capabilities, GAI has seen increasing use in
fields such as architecture and law for producing documentation and
drawings. Its ability to automate mundane, repetitive tasks is likely to
revolutionize the workplace and vastly improve productivity in the coming
years \cite{bowles_2023_generating}. As of February 2023, ChatGPT is
estimated to have grown to nearly 100 million monthly active users, with
approximately 13 million unique visitors per day \cite{hu_2023_chatgpt}.
Furthermore, with future improvements in its usefulness and
cost-effectiveness, its usage is only likely to become more sophisticated
and widespread in a multitude of industries. Predictions by
Pricewaterhouse Coopers estimate that nearly 44\% of low-skilled jobs,
including retail and sales, are likely to be taken over by automation in
the mid-2030s, whereas the number drops to only 10\% for higher-skilled
jobs. Thus, moving forward, it will be those who can automate processes
and improve their productivity with generative AI that will be most in
demand and resilient to job loss.

Along with industry, there has also been much speculation regarding
applications for GAI in education. First, one must address the purpose of
education as a whole: to improve the student’s employability and position
in the job market, and to improve their mind in terms of knowledge and
critical-thinking/problem-solving skills \cite{bowles_2023_generating}. As
previously discussed, students will need to be prepared to use GAI to
fulfill the employability aspect of these criteria; despite this, the main
controversy surrounding these tools stems from the mental improvement
aspect. On one hand, GAI could provide students with access to on-demand
personalized learning resources and feedback that suit their individual
needs; at the same time, their ability to generate completely natural,
novel text could render traditional essays obsolete, and the potential for
producing harmful misinformation simply cannot be ignored
\cite{baidooanu_2023_education}. The current policies in educational
facilities regarding generative AI, particularly at the high school level,
generally involve banning GAI outright, which discounts the potential
benefit these tools could have on student outcomes. As a result, clear
policies and guidelines must be established for the use of generative AI
in classrooms to maximize the potential benefits and mitigate the risks.

\section{Literature Review}
\subsection{Technology in Education}
In the past few decades, there has been a greater push for the
implementation of technology in education. Said efforts have proven to be
costly affairs, subject to cost overruns and delays due to a lack of
accountability \cite{christensen_2001_instruments}. Thus, to successfully
implement a given technology in classrooms, clear guidelines and rules are
needed to ensure accountability and ensure that the full benefits of the
technology are reaped. In a similar vein, to reap these benefits,
wide-scale efforts to use the technology are needed: teachers must
demonstrate willingness to use the technology; additionally, both teachers
and students must possess good technology literacy and understand the
technology \cite{moursund_2005_introduction}. The original Technology
Acceptance Model (TAM) described by Davis et al. quantifies this
relationship, theorizing that the perceived usefulness and perceived ease
of use of technology determine the users’ attitude towards said
technology, which is directly responsible for the actual use of the
technology \cite{davis_1989_user}. Thus, especially for a technology like
generative AI that has such extensive exposure in the modern world, much
of which is negative (regarding job loss and academic dishonesty),
educating users (students and educators) on the potential benefits and
limitations of GAI will be an important first step towards ensuring a
smooth adoption of the technology in classrooms. Going back to the study
by Moursund in 2005 finds that if the two criteria are met, technology can
allow for a more interactive and personalized learning experience, which
improves student engagement. This results in improvements in student
performance and learning outcomes, fostering the development of
higher-order thinking skills while also preparing students for the
increasingly technology-dependent workplace of the future.

\subsection{Applications of Generative AI in Education}
The use of generative AI in education has been the subject of significant
debate and controversy; however, it is undeniable that it does have some
strong applications in the field of education and has the potential to
bring clear benefits to students. For one, tools such as ChatGPT can be
used to provide students with around-the-clock access to personalized
learning experiences in the form of tutoring, practice questions, and
other resources that can be tailored to the student’s individual needs and
adapt based on their progress in real-time
\cite{okaiyeto_2023_generative}. As a result, students who otherwise might
not have had access to this kind of personalized, out-of-classroom
assistance (for any number of reasons, including cost and accessibility),
will now be able to utilize it to their benefit, thus furthering the cause
of equity in educational opportunities. Circling back on the concept of
equity, the study also found that individuals who used ChatGPT were much
more productive in writing tasks, proposing that this could disrupt the
academic writing field and bridge productivity gaps. Although this
argument seems to support the idea that students should be allowed to use
GAI on writing assessments to bridge performance gaps as well, the
importance of developing strong writing and communication skills
independently of these tools must be noted, as this is a highly desirable
trait in both academic and professional settings.

Building upon the idea of improving generative AI literacy, one commonly
proposed method is for the education institution itself to offer
generative AI literacy programs or workshops to help foster awareness and
mindfulness \cite{okaiyeto_2023_generative, kong_2022_evaluating}. The
former study by Kong et al. [2022] proposes that there are three aspects
to generative AI literacy: cognitive, affective, and sociocultural.
Cognitive refers to an overall understanding of generative AI and its
impacts in the real world, including its capabilities and limitations.
Affective involves the perceived usefulness of the GAI, along with its
ease of use. Finally, sociocultural involves understanding the ethical
impacts of generative AI and how to use it responsibly. By addressing all
of these aspects adequately, the study finds that students felt decidedly
more empowered to use GAI responsibly and ethically. This is significant,
as improvements in self-confidence in a particular subject have been
proven to have a definite impact on student outcomes in that subject
\cite{valentine_2004_the}. Thus, generative AI can be a useful tool in
educators’ and students’ hands to improve learning outcomes, although the
proper steps must be taken to improve GAI literacy and train stakeholders
on its use beforehand to optimize this impact.
\subsection{Limitations of Generative AI}
Despite the undeniable potential for benefit as a result of using GAI in
classrooms, it is just as important to acknowledge its limitations and its
consequences. To begin, the aforementioned ability to generate
natural-sounding text based on a prompt has been a clear point of
contention as it is rife with the potential for unchecked plagiarism on
traditional essay-based assignments and assessments
\cite{nah_2023_generative}. Due to the novel nature of this text,
traditional plagiarism detectors cannot detect AIGC, and specifically
designed AI checkers are not accurate enough to be reliably used to check
students’ work. As a result, the potential for academic dishonesty is
quite significant and there is essentially no way to detect it, which is
why the aforementioned literacy programs are so valuable, culling
potential academic dishonesty at its very root. Going back to the subject
of limitations, the combination of these factors could lead to students
being over-reliant on GAI to complete assignments and in the learning
process as a whole, which could potentially limit learning and worsen
academic performance rather than improve it. Beyond the human aspect, the
GAI models themselves have some limitations that must be addressed. As a
result of their generative nature, ensuring factual correctness and
accuracy is a particularly daunting challenge for generative AI models –
they often tend to fabricate sources or fabricate the information in the
sources \cite{zhao_2023_can}. In the hands of students, this could lead to
the spread of misinformation and if not vetted properly, an incorrect
understanding of a given topic given to the student by the GAI. More
concerningly, alongside factual inaccuracies, most GAI models are trained
off of data from the internet, so any biases or disinformation that are in
the training dataset have the potential to be transferred into the output
of the model itself \cite{kadaruddin_2023_empowering}. This could lead to
students being subject to potentially harmful disinformation and biased
content, which in turn could lead to them taking on those biases or
disinformation as a result. Many existing models, such as Bing AI, have
taken steps to address such issues by implementing content filters to
restrict the discussion of misinformation and sensitive topics and
providing citations to the model’s output so that users can more easily
verify its output to be correct

\subsection{Overview}
The existing academic discussion surrounding generative AI in its current
state has been limited to speculation and hypotheses on its applications
and potential impacts on students. The findings of these studies, such as
Okaiyeto et al. [2023] and Nah et al [2023] have theoretical backings
based on previous research and trials conducted in adjacent fields and
thus, maintain their credibility in an academic context and the reasoning
contained therein is certainly sound. However, the real world rarely
conforms to the ideas set forth based on theoretical explanations, and
thus, it is necessary to put forth objective, experimental data collected
in real-world contexts to be able to support or refute these hypotheses
with a reasonable level of certainty. This lack of experimental data is
the gap that this study seeks to address and fill to address the primary
research question: how can generative AI be used to improve educational
outcomes in high school STEM education?
\section{Methods}

\subsection{Research Design}
This research project utilizes a quasi-experimental approach to analyze
the relationship between the use of generative AI in high-school STEM
education and student learning outcomes. These outcomes will be quantified
using objective academic performance on assessments alongside the
students’ subjective evaluation of their learning and understanding of the
topic, this secondary qualitative aspect to the assessment allows for an
evaluation of students’ self-confidence in their understanding of the
subjective, which has been shown to have a clear impact on their learning
and overall academic performance in the long term
\cite{valentine_2004_the}. Despite this, the objective aspect of the study
is also necessary to provide hard data on the effectiveness of the various
treatments in comparison to the control group, which is a key component of
the gap in the body of knowledge as previously discussed in the literature
review.

Moving on to the actual structure of the study, two primary groups are
being analyzed: AP Physics 1 students and on-level physics students. The
AP Physics 1 students have been following the AP Physics 1 curriculum as
outlined by the AP College Board, whereas the on-level physics students
have been following the curriculum outlined by the Texas Education Agency
in the Texas Essential Knowledge and Skills (TEKS) for Science.
Considering that one of the main advantages of GAI use in education is the
potential for improved equity in education due to greater access to
personalized tutoring and resources, analyzing differences in outcomes
between AP Physics 1 students (who are generally more academically
inclined) to on-level Physics students will be crucial in assessing
whether there is merit to this argument of educational equity
\cite{baidooanu_2023_education, okaiyeto_2023_generative}. With regards to
the decision to use physics as the subject of focus for this experiment,
the reasoning was that physics requires a strong conceptual foundation,
which GAI is generally well-equipped to provide (whereas it tends to not
be nearly as effective in mathematically oriented fields).

Each of these groups will be divided into four experimental groups:
control, search engines (SE), generative AI (GAI), and generative AI with
literacy program (GAI-LP). The control group will have no access to any
external resources and will rely only on discussions with peers. The SE
group will have access to a search engine, namely Google since it is the
most prevalent engine by a significant margin, but no generative AI
\cite{bianchi_2023_global}. The GAI group will be provided access to a
generative AI but given no training or assistance on how to use it
effectively in the context of physics. Lastly, the GAI-LP group will
receive access to the same generative AI, but will also be provided with a
generative AI literacy program slideshow outlining the responsible and
effective use of generative AI. All students will be making use of Bing
AI. The reasoning behind the choice of Bing AI is its use of the
cutting-edge GPT-4 model by OpenAI that promises to be more versatile and
accurate in its responses \cite{openai_2023_gpt4}. Thus, it is the
strongest publicly available generative AI and more importantly, the best
suited for educational purposes.

\subsection{Sampling}
The target population of this study will be students attending Tomball
Memorial High School (located in Tomball, TX), specifically, those
currently enrolled in a physics class. This school was primarily chosen
due to its accessibility to the researcher. There will be three AP Physics
1 classes included in the sample and four on-level physics classes, and
for consistency across each group, all students in each group will have
the same instructor (one instructor for all three AP Physics classes, and
another for all for on-level physics classes. The majority of students in
this demographic are aged 16-18, with the majority being on the younger
side, and thus parental consent was required for all minors involved (and
participant consent for adults), which was obtained using a Google Form
tailored for each experimental group as a form of blinding to prevent
potential participant biases from altering the validity or accuracy of the
data collected. Furthermore, to ensure the anonymity of the participants
and prevent bias during the data analysis phase, all participants were
provided with an experimental ID following this format: XX-P-NN, where XX
represents the teacher code to differentiate between AP and on-level
physics students, P represents the period number of the class, and NN
represents the roll of the students. These experimental IDs were
distributed to the students by the teachers with no involvement from the
researcher other than providing the format of the IDs to ensure that the
researcher is unable to link the experimental IDs to individual students.

\subsection{Procedure}
Initially, the consent forms were generated and given to the teachers as
early as possible to ensure that they would have time to distribute the
survey and obtain the maximum number of participants in the experiment. In
the end, the total participant breakdown was as follows: 170 total
participants, with 90 in AP Physics and 80 in on-level physics. In terms
of the topics chosen, for AP Physics 1, the early sections of momentum and
impulse were chosen because they are relatively concept-heavy relative to
the rest of the AP Physics 1 curriculum, which is a strength area of GAI
as previously discussed. For on-level physics, electricity and circuits
were chosen as they too require a strong conceptual understanding of the
subject matter. Ideally, to have the most fair and accurate comparison,
both AP Physics 1 and on-level physics would use the same topic; however,
due to the curricula of each, this was infeasible as having both groups
working on the same topic at the same time would require a disruption of
the standard class schedules, which would violate the ethical guidelines
of the Tomball ISD Institutional Review Board under whose purview this
study is being conducted. After a sufficient number of responses to the
informed consent was gathered to have a sufficiently high sample size for
every group, the students selected to take part in the GAI-LP group were
provided with the GAI training program slides to review and study before
the experiment.

The overall structure of the study involves a standard pre-test,
treatment, and post-test format, which is quite common in the fields of
education as it enables the researcher to control for prior knowledge of
the subject to be able to measure the actual change in student performance
\cite{tshering_2022_use, yunzal_2020_effect, hung_2021_unbundling}. As
previously discussed, each of the experimental groups will receive a
different treatment; however, before this, all of the groups will receive
a lecture from their respective instructors on the topic. The purpose of
the inclusion of this lecture aspect is to mimic a traditional
classroom/educational experience for the students, albeit on a shortened
time scale. The pre-test analyzes their understanding of the subject area
at the beginning of a given lesson, then they receive a lecture that is
intended to be analogous to the classroom instruction they would normally
receive from the teacher. The actual treatments (control, SE, GAI, GAI-LP)
are intended to represent the students’ study techniques and tools outside
of the classroom; of course, it should be noted that generally, students
tend to employ a mixture of these techniques but for this experiment and
to be able to differentiate the individual impacts of each treatment, they
are kept separate in this study. Finally, the post-test is intended to
assess students' understanding of the subject area after the lecture and
study time and as a result, is representative of their potential
performance on a test or exam. Finally, the students will also be asked to
take a post-survey to gather subjective, qualitative data on their
assessment of their performance and learning throughout the experiment.
This kind of instrument is commonly used in studies in the field of
education as students’ self-confidence in their learning and understanding
of a topic has a strong impact on their overall performance and outcomes
\cite{valentine_2004_the, hung_2021_unbundling}. The informed consent,
pre-test, post-test, and survey were administered via Google Forms to
ensure that all students could easily access them, along with the ability
to send participant responses directly to a Google Sheet where they can
then be compiled and analyzed efficiently. Concerning the slideshows for
the lecture and GAI literacy program, these were created and distributed
via Google Slides once again due to the ease of access for students and
ease of distribution for the researcher and teachers.

Class periods at Tomball Memorial High School are 45 minutes in length,
considering the time taken by students to settle down and pack up at the
beginning and end of class respectively, this was shortened to an
effective length of 40 minutes. To ensure that the students had enough
time to adequately and fully complete each element of the experiment, an
overall length of two class periods was chosen (which means a total length
of two days for the experiment as a whole). Initially, the pre-test would
be administered for 15 minutes, followed by a twenty-minute lecture given
by the teacher on the selected topic. After this point, students would be
instructed to avoid discussing or reviewing the topic before the next
class period to avoid tampering with their understanding of the material
before the post-test; however, there is no way to enforce such a policy
effectively and thus its efficacy is somewhat limited. On the next day,
students would begin with a 20-minute study period using their respective
treatment (control, search engine, or generative AI). After this, they
will proceed to the 15-minute post-test, followed by the survey that they
will have until the end of the period to complete to ensure that the
maximum number of students submit the survey. During the pre- and
post-tests, students will not be allowed to discuss amongst themselves or
use any external resources to ensure their results are accurate to their
actual understanding of the topic.

\subsection{Limitations}
The majority of generative AI products on the market right now are
general-purpose products whose primary intention is either recreational
use or to serve as a testbed for the GAI models they use. The result of
this is a distinct lack of factual accuracy in the generative AI, either
through the fabrication of sources or the content within them
\cite{zhao_2023_can}. Further, these tools are also not specifically
designed for educational use so their responses are not specifically
tailored to improve student learning. Instead, they tend to output direct
answers to questions that could lead to concerns regarding academic
dishonesty as a result of the use of generative AI. Khanmigo, a GPT-4
based generative AI tool designed by Khan Academy, hopes to solve these
issues by creating a generative AI model that focuses on leading students
to solutions rather than giving them the solutions outright to improve
their learning outcomes overall \cite{khanmigo_2023}. However, this tool
is currently in a closed beta, requiring a monetary donation to access; as
a result, it was infeasible to use for this project due to the lack of
funding. Despite this, Bing AI was chosen as its replacement since it is
openly available and based on the same GPT-4 model and as a result should
perform fairly similarly given a similar prompt.

Moving on, one of the key benefits of the use of most digital learning
aids and generative AI in particular (due to the heightened level of
personalization) is the increase in student engagement in the learning
process, and as a result, an increase in the students’ interest in the
subject \cite{kadiyala_2000_a} \cite{youssef_2022_ict}. These benefits
become much more pronounced with long-term use of the tools, and so the
ideal time frame for a study such as this one would be over the course of
an entire academic semester or academic year. However, as previously
mentioned, this is beyond the limitations of what is allowable under the
Tomball ISD IRB regulations and would be infeasible to accomplish as part
of the AP Research course. Thus, as previously discussed, a condensed
version of the overall learning process has been chosen for the
methodology of this study which will span only two school days. While this
may not necessarily be enough to fully determine the effects of the
various treatments on student performance, it will provide a reasonable
short-term estimate of student improvements that can be used to gauge what
the effectiveness of those treatments might be over the long term.

Lastly, because of the nature of this experiment – having to be split over
two school days – there is a gap between the lecture and post-test phase
during which the student’s actions cannot be controlled, and thus, there
is no certainty that the students did not study or review the topic on
their own time or even discuss the experiment with their classmates who
may also be participants in the experiment, which could lead to some
variation in the results that is not due to the actual treatments being
applied. To combat this, students will be informed that the pre- and
post-tests will not be taken for a grade. This will prevent students from
attempting to gain an advantage in the assessments for fear that they will
impact their academic record. Despite this, that same effect could lead to
students not being motivated to put effort into the assessments and
surveys; however, this is a necessary step that must be taken to ensure
the validity of the experiment and its resulting data.

\section{Results and Discussion}
\subsection{Data Overview}
\begin{table}[h]
    \centering
    \caption{Quantitative Data Overview}
    \label{tab:quantOverview}
    {
        \begin{tabular}{lrrr}
            \toprule
            \multicolumn{1}{c}{} & \multicolumn{2}{c}{Section} & \multicolumn{1}{c}{}         \\
            \cline{2-3}
            Experimental Group   & AP Physics                  & On-Level Physics     & Total \\
            \cmidrule[0.4pt]{1-4}
            SE                   & $33$                        & $19$                 & $52$  \\
            Control              & $28$                        & $23$                 & $51$  \\
            GAI                  & $17$                        & $17$                 & $34$  \\
            GAI-LP               & $13$                        & $21$                 & $34$  \\
            Total                & $91$                        & $80$                 & $171$ \\
            \bottomrule
        \end{tabular}
    }
\end{table}
There were a total of 171 participants in the quantitative aspect of the study, with 91 from AP Physics and 80 from on-level physics. The distribution of students between each of the samples was fairly even, although since the GAI and GAI-LP were from the same class period, their size is comparatively smaller, although this should not affect the validity of the results in any significant manner. For the purposes of data analysis, the experimental groups were compared against other groups from the same section since the quantity of responses was sufficiently large.

\begin{table}[H]
    \centering
    \caption{Qualitative Data Overview}
    \label{tab:qualOverview}
    {
        \begin{tabular}{lrrr}
            \toprule
            \multicolumn{1}{c}{} & \multicolumn{2}{c}{Section} & \multicolumn{1}{c}{}         \\
            \cline{2-3}
            Experimental Group   & AP Physics                  & On-Level Physics     & Total \\
            \cmidrule[0.4pt]{1-4}
            SE                   & $2$                         & $2$                  & $4$   \\
            Control              & $6$                         & $9$                  & $15$  \\
            GAI                  & $3$                         & $18$                 & $21$  \\
            GAI-LP               & $2$                         & $18$                 & $20$  \\
            Total                & $13$                        & $47$                 & $60$  \\
            \bottomrule
        \end{tabular}
    }
\end{table}
Moving on to the qualitative aspect, i.e. the survey, there was only a total of 60 responses, split between 47 in on-level physics and only 13 in AP Physics. Considering the limited sample size in AP Physics, conducting individual analyses on each section would not have been feasible; thus, considering this along with the fact that the tools used in both sections were the same, a combined analysis of the experimental groups between both sections was conducted (eg. Control would consist of the AP and on-level physics sections). Furthermore, considering that the GAI group only had 5 responses even after this, the GAI and GAI-LP groups were merged to improve the reliability of the analysis.

\subsection{Assessment Results}
As previously mentioned, the pre- and post-tests were distributed to the
participants as Google Forms, the results of which were exported into a
Google Sheet and collated based on experimental ID. Considering that the
focus of this study is on the differences between the pre- and post-tests,
the differences between the two assessments were then calculated for each
participant and exported into the JASP data analysis software for further
statistical analysis.

First, descriptive statistics were calculated for each section to
determine if there were any overarching trends in the data prior to
conducting an in-depth analysis to provide a greater understanding of the
results and contextualize them.

\begin{table}[h]
    \centering
    \caption{AP Physics Score Difference Descriptives}
    \label{tab:AP-Descriptives-Difference}
    {
        \begin{tabular}{lrrrrr}
            \toprule
            Experimental Group & N    & Mean     & SD      & SE      & Coefficient of variation \\
            \cmidrule[0.4pt]{1-6}
            Control            & $26$ & $-0.346$ & $1.355$ & $0.266$ & $-3.914$                 \\
            SE                 & $31$ & $0.484$  & $1.387$ & $0.249$ & $2.867$                  \\
            GAI-LP             & $14$ & $0.500$  & $1.225$ & $0.327$ & $2.449$                  \\
            GAI                & $14$ & $-0.857$ & $1.791$ & $0.479$ & $-2.090$                 \\
            \bottomrule
        \end{tabular}
    }
\end{table}

\begin{table}[h]
    \centering
    \caption{On-Level Physics Score Difference Descriptives}
    \label{tab:OL-Descriptives-Difference}
    {
        \begin{tabular}{lrrrrr}
            \toprule
            Experimental Group & N    & Mean    & SD      & SE      & Coefficient of variation \\
            \cmidrule[0.4pt]{1-6}
            Control            & $21$ & $0.905$ & $1.700$ & $0.371$ & $1.879$                  \\
            SE                 & $16$ & $1.563$ & $2.250$ & $0.563$ & $1.440$                  \\
            GAI                & $14$ & $0.714$ & $2.199$ & $0.588$ & $3.078$                  \\
            GAI-LP             & $21$ & $1.190$ & $1.569$ & $0.342$ & $1.318$                  \\
            \bottomrule
        \end{tabular}
    }
\end{table}

In both sections, the GAI group performed worse than the control,
indicating that the GAI alone could actually be detrimental to students.
The SE group performed better than the control in both sections as well,
taking these two together, it appears as if an untrained individual could
experience more academic success using a search engine as opposed to a
generative AI. However, it is also quite clear that in both sections, the
GAI-LP group performed significantly better than the GAI group,
demonstrating that the Literacy Program had a significant impact on the
students’ performance.

For a more in-depth analysis of the data, a one-way ANOVA test was
conducted for the differences in test score difference across experimental
groups for each section. Despite the general trends in the descriptive
statistics, it was important to run these tests to determine whether those
trends had any statistical merit and whether there was any statistically
significant difference between the experimental groups. Additionally, as
per convention, an alpha of 0.05 was chosen as the significance level,
meaning that any p-value below that level would indicate that the results
of the test were significant. The ANOVA test for on-level physics
[\ref{tab:ol-anova-difference}] returned a p-value of 0.619 (p > 0.05),
which indicates that there is not enough evidence to support any
significant statistical differences between the control groups.

\begin{figure}[H]
    \centering
    \begin{subfigure}{0.4\textwidth}
        \includegraphics[width=0.9\linewidth]{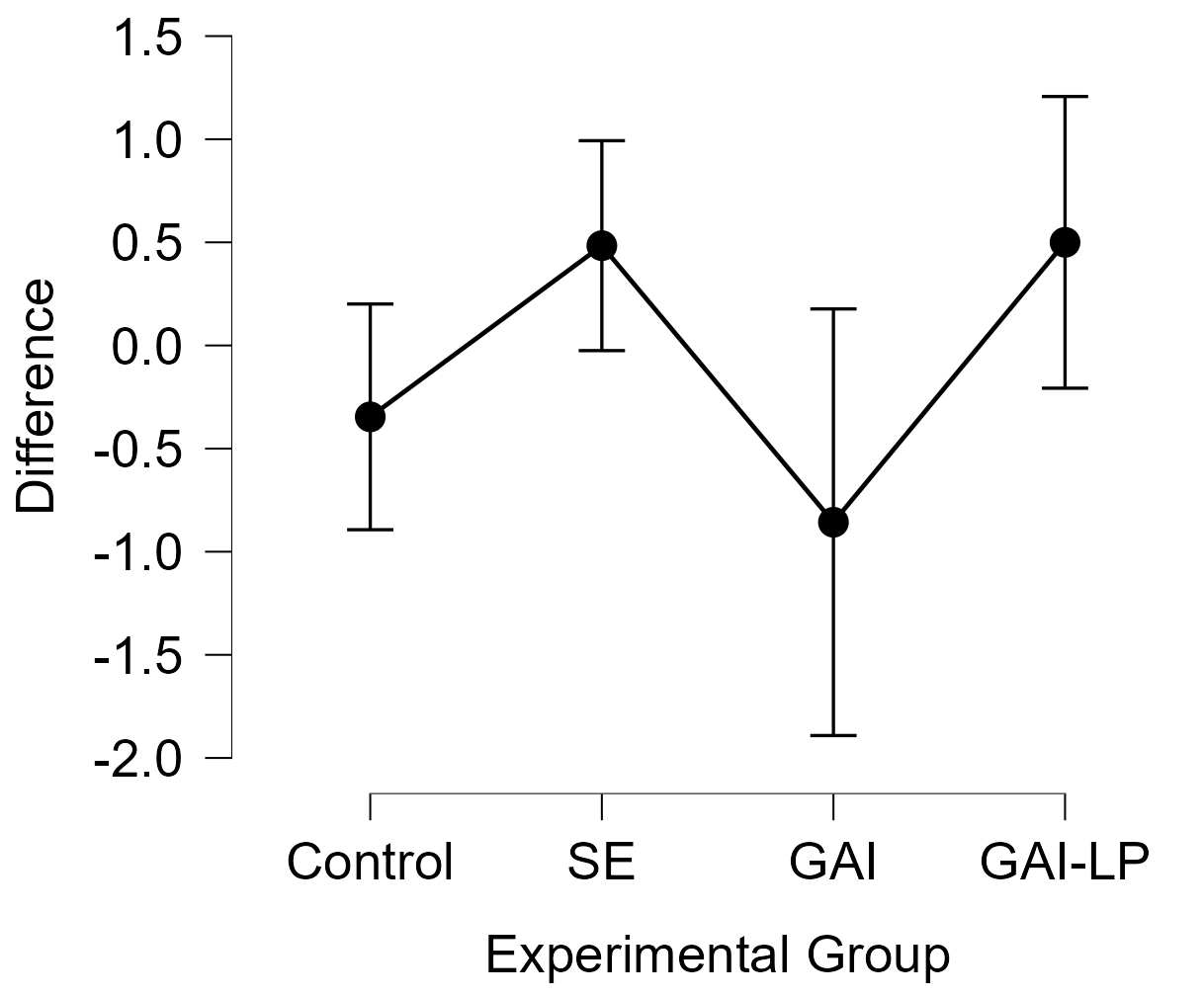}
        \caption{AP Physics}
        \label{fig:ap-descriptive}
    \end{subfigure}
    \begin{subfigure}{0.4\textwidth}
        \includegraphics[width=0.9\linewidth]{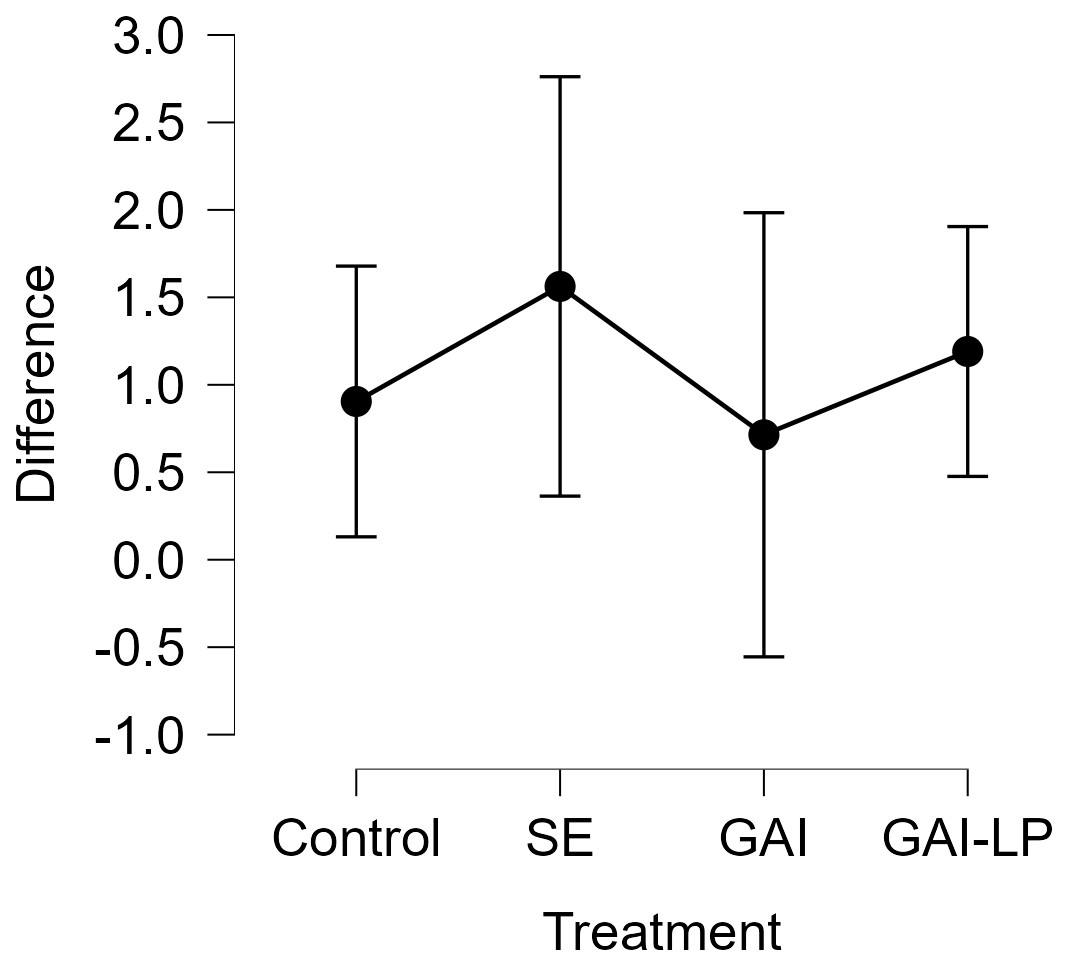}
        \caption{On-Level Physics}
        \label{fig:ol-descriptive}
    \end{subfigure}
    \caption{Descriptives Plots}
    \label{fig:descriptives}
\end{figure}

However, the test for AP Physics [\ref{tab:ap-anova-Difference}] returned
a p-value of 0.011 ($p < 0.05$), since this is below the significance
level of the test, it can be concluded that there is a statistically
significant difference between the experimental groups in this section.
Based on the interval plot, the most significant differences appeared to
be between the SE group ($M = 0.484$) and the GAI group ($M = -0.857$),
the GAI group ($M = -0.857$) and the GAI-LP group ($M = 0.50$). In order
to confirm this, independent-sample t-tests were performed between those
two pairs. The t-test for the SE and GAI groups returned a p-value of
0.0062, whereas the test between the GAI and GAI-LP groups returned a
p-value of 0.0256. In both cases, the p-value was less than the
significance level which shows that a statistically significant difference
is present between the pairs.

\subsection{Survey Results}
\begin{figure}[H]
    \centering
    \begin{subfigure}{.3\textwidth}
        \includegraphics[width=0.9\linewidth]{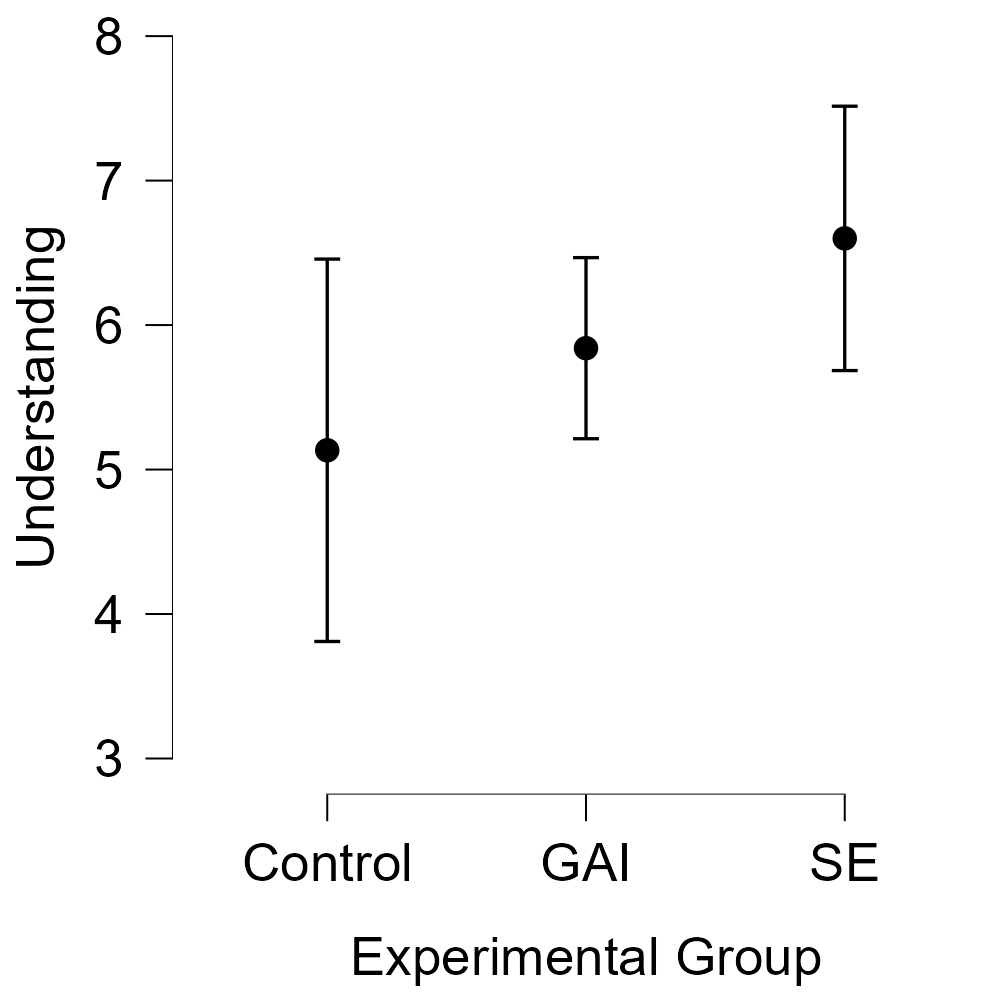}
        \label{fig:understanding-int}
    \end{subfigure}
    \begin{subfigure}{.3\textwidth}
        \includegraphics[width=0.9\linewidth]{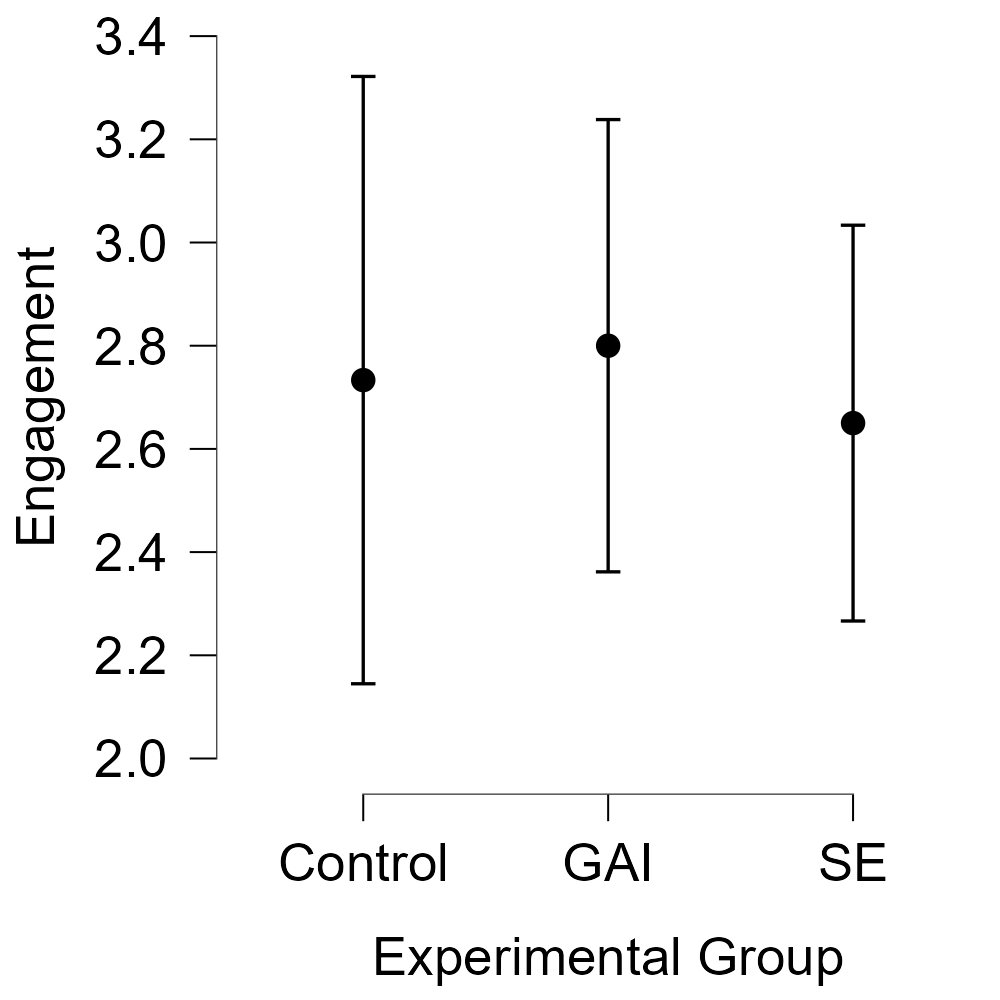}
        \label{fig:engagement-int}
    \end{subfigure}
    \begin{subfigure}{.3\textwidth}
        \includegraphics[width=0.9\linewidth]{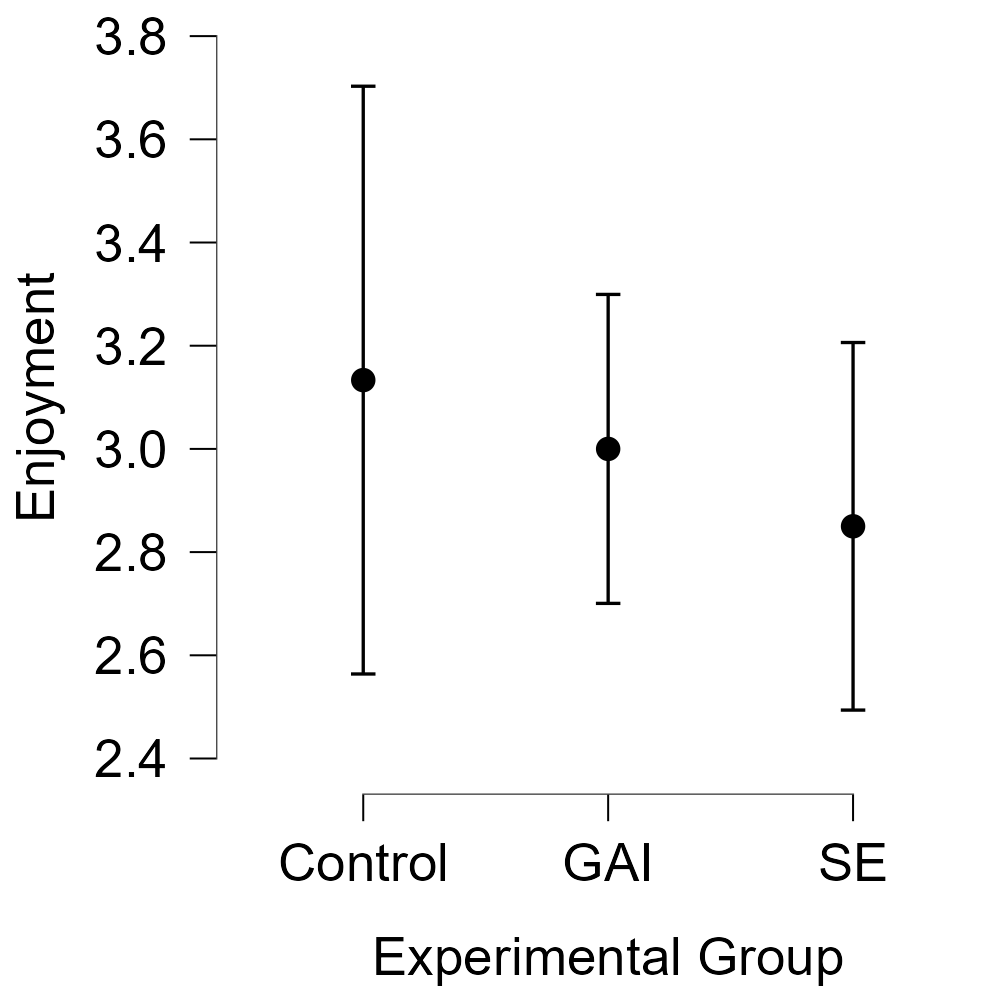}
        \label{fig:enjoyment-int}
    \end{subfigure}
    \caption{Interval Plots of Survey Responses}
    \label{fig:survey-interval-plots}
\end{figure}

Similar to the assessment results, the survey results were also compiled
in a Google Sheet and then directly exported into JASP for further
analysis, beginning with a descriptives table to identify any potential
trends that might be present in the data. Based on this table, the
majority of variables collected were nearly identical across the
experimental groups; however, three variables in particular appeared to
show some degree of variance, and as a result, they were chosen to proceed
with a further, more in-depth analysis: understanding of the topic,
engagement during the learning process, and enjoyment of the learning
process. Considering the general consensus in the body of knowledge
discussed in the literature review, these were some key predicted areas in
which GAI would have an impact on students. To begin this in-depth
analysis, an interval plot with 95\% confidence intervals was plotted as
shown below [\ref{fig:survey-interval-plots}].

Based on these intervals, it is clear that there is significant overlap
between the confidence intervals and thus, it is unlikely that there is
any statistically significant difference across the experimental groups
for these variables. However, looking at the means alone, some trends can
be observed. Starting off with understanding, which is undeniably
important for any student’s academic success, the SE group averaged the
highest understanding of the three groups, with GAI coming just after, and
the control group coming last. On the other hand, considering engagement,
the GAI group reported having slightly higher engagement than all the
other groups, although this difference is marginal at best. In enjoyment,
the GAI group ranked slightly below the control, and slightly above the SE
group, although again the overlap is too large to draw any conclusions.

In order to confirm this, a MANOVA test [\ref{tab:mANOVA:PillaiTest}] was
run with understanding, confidence, and engagement as the dependent
variables. Of the various types of MANOVA, Pillai’s Trace and Wilke’s
Lambda are the most appropriate candidates for this case as the sample
sizes are relatively low. Although the conditions for homogeneity and
normality of variances are met, the sample sizes of the experimental
groups are rather unequal, and thus Pillai’s Trace is most suitable as it
is comparatively resilient to a violation of this condition. After running
this test, the calculated p-value was 0.115 ($Trace_{Pillai} = 0.136$),
this is significantly greater than the significance level at 0.05, so as
expected, there is no statistically significant difference between the
experimental groups.

\begin{table}[H]
    \centering
    \caption{MANOVA of Survey Results: Pillai Test}
    \label{tab:mANOVA:PillaiTest}
    {
        \begin{tabular}{lrrrrrr}
            \toprule
            Cases              & df   & Approx. F & Trace$_{Pillai}$ & Num df & Den df   & p       \\
            \cmidrule[0.4pt]{1-7}
            Experimental Group & $1$  & $2.107$   & $0.136$          & $3$    & $40.000$ & $0.115$ \\
            Residuals          & $42$ & $$        & $$               & $$     & $$       & $ $     \\
            \bottomrule
        \end{tabular}
    }
\end{table}

Finally, distribution plots of the GAI-LP respondents were also plotted to
ascertain how much of an impact they felt the GAI-LP had on their
performance and understanding of GAI. Considering the impact on their
assessment scores, it was predicted that this would translate into their
survey responses, but that did not seem to be the case. The majority of
respondents stated the overall impact was a 3 on a scale from 1-5, which
represents no impact, indicating that the perceived effectiveness of the
literacy program was quite low whereas its observed effectiveness is much
higher.

\begin{figure}[H]
    \centering
    \begin{subfigure}{0.4\textwidth}
        \includegraphics[width=0.9\linewidth]{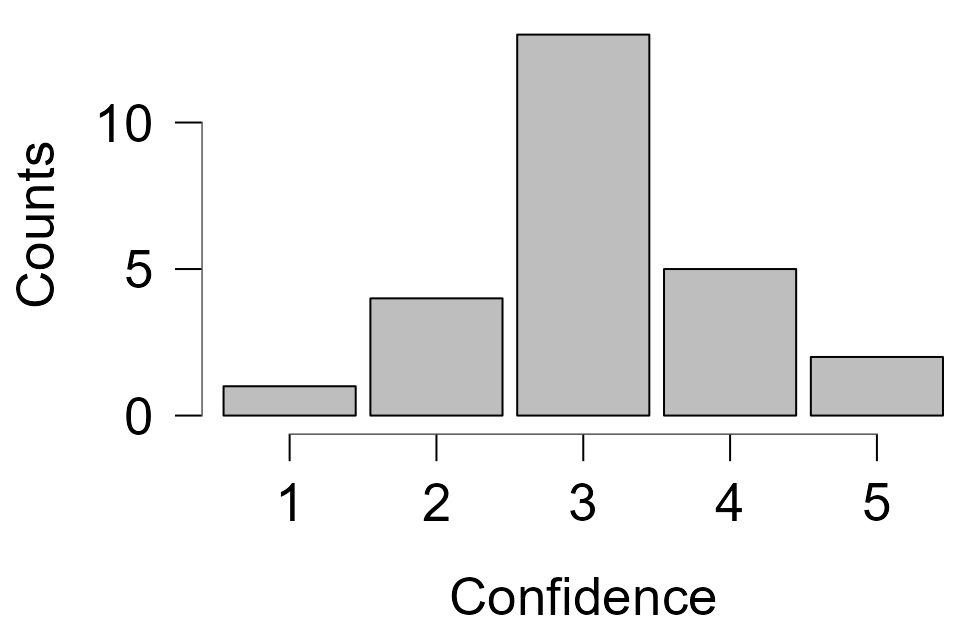}
        \label{fig:gailp-confidence}
    \end{subfigure}
    \begin{subfigure}{0.4\textwidth}
        \includegraphics[width=0.9\linewidth]{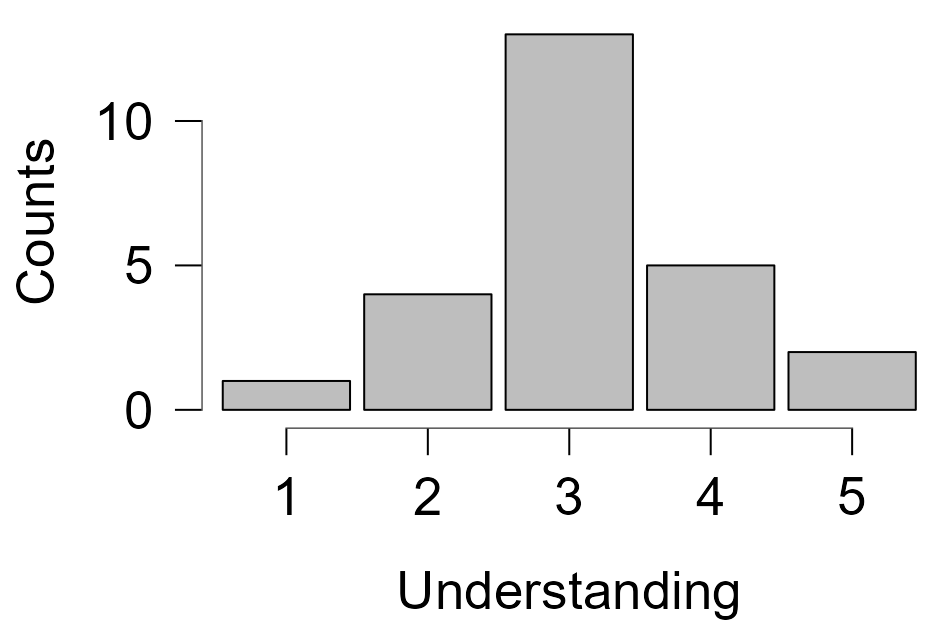}
        \label{fig:gailp-understanding}
    \end{subfigure}
    \caption{Distribution Plots for GAI-LP}
    \label{fig:gailp-plots}
\end{figure}

\subsection{Analysis}
With regards to the quantitative data from the assessments, there was some
significant differences between the experimental groups as anticipated.
One trend that was rather significant was the GAI group not only scoring
significantly worse than the GAI-LP group across both sections, but also
performing worse than the SE group and even the control group. Researchers
in Kong et al., 2022 found that students who took part in the GAI Literacy
Program felt more confident about their ability to effectively use GAI and
the results from the assessments indicate that there is not only an
improvement in the students’ confidence to use GAI but also in their
objective academic performance. Even more significant is the fact that
students who used GAI without the literacy program actually performed
worse than the control group, indicating that the use of GAI without a
strong understanding of how to use it ethically and responsibly can
actually be detrimental to students’ academic performance, making it even
more crucial that schools and districts take this into consideration and
educate students on ethical and effective GAI use.

Despite this clearly significant improvement with the literacy program,
these results do not translate into the survey results where the majority
of respondents stated that they felt the GAI literacy program had no
impact on their performance. Initially, the existence of a confounding
variable such as class scheduling leading to higher performing students
being placed in a particular class period. However, the study was
specifically designed to counteract this kind of situation through the use
of the pre- and post-test format, which accounts for previous knowledge
and only looks into the difference between the student’s initial and final
performance. Furthermore, in the AP Physics section, the GAI-LP and GAI
groups were from the same class period, so any differences in their test
scores could not be explained through scheduling, and this along with the
consistency of the trend of the GAI-LP group vastly out-scoring the GAI
group makes the likelihood of this kind of a confounding variable causing
this variation to be too small to be of any significance. Thus, for the
purposes of this study, the existence of such variables shall be
considered to be outside of its purview, although further investigation
could be beneficial to the body of knowledge.

Although there was statistically significant improvement between the GAI
and GAI-LP groups in AP Physics, it still stands that in both sections,
there was no significant difference between the control group and any one
of the testing groups, including the GAI-LP group. The lack of variation
between the control group and any of the experimental groups indicates an
overall lack of improvement or even change in academic performance across
the board with the application of the treatments. This is not consistent
with the general consensus from prior studies in the body of knowledge
such as Bowles \& Kruger \cite{bowles_2023_generating} or Baidoo-Anu \&
Ansah \cite{baidooanu_2023_education} which predicted that the use of GAI
would have a much more significant effect on students performance than
what was observed in the assessments and a vastly more significant effect
on the students’ engagement in and enjoyment of the learning process than
was observed. In fact, studies such as Youssef et al.
\cite{youssef_2022_ict} and Kadiyala \& Crynes \cite{kadiyala_2000_a} have
found that the use of technology in and of itself, GAI notwithstanding,
has a significant impact on student engagement through the increased
interactiveness of the learning experience. Thus, this lack of observed
impact in the survey and in the assessment results does appear to be quite
anomalous. One potential explanation for this could be the comparatively
short-term nature of this study. The twenty minutes allotted to the
students to utilize the various treatments may not have been enough for
them to fully reap its benefits or become engaged with it, resulting in
the observed results.

\section{Conclusions and Future Directions}
Considering the data results, it is clear that the use of GAI will be an
important subject that must be actively addressed by educators and school
districts if they are to adequately prepare their students for not only
the future of higher education, but the future of the workplace and
society as a whole. GAI is becoming increasingly prevalent in all aspects
of society, and it will be those who understand how to use it effectively
to better themselves who will be most sought after. On that note, GAI is
already seeing widespread use in education despite educators’ best efforts
to eliminate it, and as per the assessment results, it can definitely be
detrimental to students performance if they are left to use it without
being educated as to how to do so responsibly and ethically, in a way that
benefits them and their education. However, it is clear that with a
generative AI literacy program implemented into education, students will
be able to take full use of the technology at their hands to improve their
learning and become more engaged in the process.

As previously discussed, the short-term nature of this study means that
although it is able to provide an inkling of the impact GAI can have on
students’ education, it cannot provide the full picture. Thus, a more
long-term study over the course of a semester or even a full academic year
tracking the academic performance of students either through their grades
or through periodic check-ins would be able to address this much more
thoroughly and provide a more complete and insightful understanding of how
GAI might impact students in the real world. Furthermore, this study
focused primarily on physics, as it is a challenging subject where there
is a lot of room for improvement, however, to broaden its reach, a study
looking into various areas of study between the humanities, language arts,
mathematics, and science could prove to be beneficial in assessing how GAI
impacts disciplines with different ways of thinking and conventions that
might lend themselves more or possibly less to the effective use of GAI.
Ideally, both of the aforementioned studies would be combined to best fill
the gap in the body of knowledge; however, such an effort would require a
large amount of planning and coordination, not to mention a wide-scale
effort across departments to be effective and thus its feasibility would
be in question. Finally, one key factor that could not be addressed in
this study is the use of an education-specific GAI model. Khanmigo was the
initial candidate for this study as it was developed specifically through
an educational point-of-view and was geared towards leading students
towards the answers rather than providing them outright, and thus would be
more beneficial to students learning. Due to a lack of funding, Khanmigo
could not be utilized in this study; however, a future study focusing on
the difference between it and a general purpose GAI could show the
benefits of having an educational GAI in the classroom and thus encourage
more educators to implement such technologies into the classroom due to
its inbuilt filters against academic dishonesty, which is still a major
concern with the aforementioned general purpose GAI models such as Bing AI
(GPT-4 based) which was used in this study.

\newpage
\bibliographystyle{acm}
\bibliography{bibliography}

\appendix
\section{Definitions}
The following definitions are not universal, but are generally understood
in the literature
\subsubsection{\textbf{Generative AI (GAI)}}
    Generative AI refers to an Artifical Intelligence that is capable of generative text, images, videos, or other data using generative models, often in response to a prompt

\subsubsection{\textbf{Artifical Intelligence (AI)}}
    Refers to a computer system able to perform tasks that generally require human intelligence (visual perception, speech recognition, decision-making, etc.)

\subsubsection{\textbf{GAI Literacy Program (GAI-LP)}}
    Refers to an educational program designed to provide awareness regarding the working of GAI, its uses, and its limitations in order to enable more responsible and ethical use of GAI.

\subsubsection{\textbf{Generative Pre-Trained Transformer (GPT)}}
    A type of LLM that is trained on large datasets of unlabeled text and is able to generate novel human-like (aka natural) text.

\subsubsection{\textbf{Large Language Model (LLM)}}
    A language model capable of generating language and achieving other natural language processing tasks such as classification (i.e. interpretation of human language).

    \section{Data Collection Forms}
    (*) indicates a required question
    \subsection{Informed Consent}
\subsubsection{Control Group} \hfill \\
    {\textbf{Email (*): } \_\_\_\_} \\
    \subparagraph{\textbf{INFORMED CONSENT FOR HUMAN PARTICIPATION}} \hfill \\
    {\textbf{Purpose}}

    My name is <researcher> and I am an AP Research Capstone student. In
    this research study, I am interested in looking at how the use of
    generative AI in high school classrooms, particularly with regards to
    STEM subjects (more specifically, on-level physics and AP Physics 1),
    will impact students' educational outcomes. I will be using an
    experimental method, giving students pre- and post-tests along with
    surveys to determine the impact of various learning aides on their
    performance and understanding of a given topic. \\ \hfill

    \subparagraph{\textbf{What can I expect?}} \hfill

    Your participation will involve a pre-test, which should take
    approximately 10 minutes. After this, the study itself will span one
    full class period (45 minutes). You will receive a short lecture (~15
    minutes) followed by 20 minutes to study as you please (without any
    kind of digital devices or internet access) and then a post-test (~10
    minutes) to assess any changes in your performance. After this, you
    will also be requested to fill out a short survey regarding how you
    felt about the learning experience. Your email will be collected
    throughout these forms but will not be used for any purposes and will
    not be shared with anyone but the researcher.\newline Your
    participation is entirely voluntary and if you choose to participate,
    you can leave the study at any time with no penalties. If you choose
    to leave, your results will not be considered. There will be no grades
    associated with this study, therefore it will have no impact on your
    academic record. The data collected from your participation will be
    recorded, but all personal data and identifying information will be
    kept fully confidential and will not be shared with anyone other than
    the researcher. \\

    \subparagraph{\textbf{Risks and Discomforts}} \hfill

    There is minimal/no risk of discomfort associated with this study.\\

    \subparagraph{\textbf{Benefits}} \hfill

    The information collected from this study may help improve the current
    understanding of how generative AI can be used in the education
    system, and as a result help develop better curricula that prepare
    students better for their futures. \\

    \subparagraph{\textbf{Contact Information}} \hfill

    If you have any questions, you can reach out to me at <researcher
    email>, or my research advisor <advisor email> \\

    \subparagraph{\textbf{Parental Minor Consent (if applicable)}} \hfill

    As a parent or guardian of a minor child, participation in this study
    requires parental consent. You are free to talk to anyone you trust
    about this study and take time to reflect whether you wish your child
    to participate or not. Please read the information about the study
    listed above carefully. \\

    \subparagraph{\textbf{2.}} \textbf{I am a student in an on-level physics or AP Physics 1 class at <name>
        High School or <name> High School, and I (and my parent/legal guardian) have
        read this form and have been able to ask questions of the primary investigator
        and/or discuss my participation with someone I trust. I (and my parent/legal
        guardian) understand that I can ask additional questions at any time during this
        research study and am free to withdraw from participation at any time.}\newline

    \subparagraph{(Mark only one oval)} (*)\hfill \\
    \_\_\_ I am above the age of 18, and I consent to my participation in this survey.\newline
    \_\_\_ I am above the age of 18, and I do not consent to my participation in this survey, and
    I wish to opt out of the survey.\newline
    \_\_\_ I am under the age of a 18, and myself and my parents or guardians have consented
    to my participation in this survey.\newline
    \_\_\_ I am under the age of 18, and myself or my parents or guardian have not consented to my participation in this survey and I wish to opt out of this survey.\newline

    \subparagraph{\textbf{3.}} \textbf{Enter your Participant ID (provided by your teacher). This should follow the format
        XX-0-00 (*)}\\ \vspace{3mm}
    \_\_\_\_\_\_\_\_\_\_\_\_

\subsubsection{SE Group} \hfill \\
    {\textbf{Email (*): } \_\_\_\_} \\
    \subparagraph{\textbf{INFORMED CONSENT FOR HUMAN PARTICIPATION}} \hfill \\
    {\textbf{Purpose}}

    My name is <researcher> and I am an AP Research Capstone student. In
    this research study, I am interested in looking at how the use of
    generative AI in high school classrooms, particularly with regards to
    STEM subjects (more specifically, on-level physics and AP Physics 1),
    will impact students' educational outcomes. I will be using an
    experimental method, giving students pre- and post-tests along with
    surveys to determine the impact of various learning aides on their
    performance and understanding of a given topic. \\ \hfill

    \subparagraph{\textbf{What can I expect?}} \hfill

    Your participation will involve a pre-test, which should take
    approximately 10 minutes. After this, the study itself will span one
    full class period (45 minutes). You will receive a short lecture (~15
    minutes). After this, you will be given 20 minutes to study this topic
    using a search engine of your choice (you may not use any other tools)
    followed by a post-test (~10 minutes) to assess any changes in your
    performance. After this, you will also be requested to fill out a
    short survey regarding how you felt about the learning experience.
    Your email will be collected throughout these forms but will not be
    used for any purposes and will not be shared with anyone but the
    researcher. \newline Your participation is entirely voluntary and if
    you choose to participate, you can leave the study at any time with no
    penalties. If you choose to leave, your results will not be
    considered. There will be no grades associated with this study,
    therefore it will have no impact on your academic record. The data
    collected from your participation will be recorded, but all personal
    data and identifying information will be kept fully confidential and
    will not be shared with anyone other than the researcher. \\

    \subparagraph{\textbf{Risks and Discomforts}} \hfill

    There is minimal/no risk of discomfort associated with this study.\\

    \subparagraph{\textbf{Benefits}} \hfill

    The information collected from this study may help improve the current
    understanding of how generative AI can be used in the education
    system, and as a result help develop better curricula that prepare
    students better for their futures. \\

    \subparagraph{\textbf{Contact Information}} \hfill

    If you have any questions, you can reach out to me at <researcher
    email>, or my research advisor <advisor email> \\

    \subparagraph{\textbf{Parental Minor Consent (if applicable)}} \hfill

    As a parent or guardian of a minor child, participation in this study
    requires parental consent. You are free to talk to anyone you trust
    about this study and take time to reflect whether you wish your child
    to participate or not. Please read the information about the study
    listed above carefully. \\

    \subparagraph{\textbf{2.}} \textbf{I am a student in an on-level physics or AP Physics 1 class at <name>
        High School or <name> High School, and I (and my parent/legal guardian) have
        read this form and have been able to ask questions of the primary investigator
        and/or discuss my participation with someone I trust. I (and my parent/legal
        guardian) understand that I can ask additional questions at any time during this
        research study and am free to withdraw from participation at any time.}\newline

    \subparagraph{(Mark only one oval)} (*)\hfill \\
    \_\_\_ I am above the age of 18, and I consent to my participation in this survey.\newline
    \_\_\_ I am above the age of 18, and I do not consent to my participation in this survey, and
    I wish to opt out of the survey.\newline
    \_\_\_ I am under the age of a 18, and myself and my parents or guardians have consented
    to my participation in this survey.\newline
    \_\_\_ I am under the age of 18, and myself or my parents or guardian have not consented to my participation in this survey and I wish to opt out of this survey.\newline

    \subparagraph{\textbf{3.}} \textbf{Enter your Participant ID (provided by your teacher). This should follow the format
        XX-0-00 (*)}\\ \vspace{3mm}
    \_\_\_\_\_\_\_\_\_\_\_\_

\subsubsection{GAI Group} \hfill \\
    {\textbf{Email (*): } \_\_\_\_} \\
    \subparagraph{\textbf{INFORMED CONSENT FOR HUMAN PARTICIPATION}} \hfill \\
    {\textbf{Purpose}}

    My name is <researcher> and I am an AP Research Capstone student. In
    this research study, I am interested in looking at how the use of
    generative AI in high school classrooms, particularly with regards to
    STEM subjects (more specifically, on-level physics and AP Physics 1),
    will impact students' educational outcomes. I will be using an
    experimental method, giving students pre- and post-tests along with
    surveys to determine the impact of various learning aides on their
    performance and understanding of a given topic. \\ \hfill

    \subparagraph{\textbf{What can I expect?}} \hfill

    Your participation will involve a pre-test, which should take
    approximately 10 minutes. After this, the study itself will span one
    full class period (45 minutes). You will receive a short lecture (~15
    minutes). After this, you will be given 20 minutes to study this topic
    using a generative AI, followed by a post-test (~10 minutes) to assess
    any changes in your performance. After this, you will also be
    requested to fill out a short survey regarding how you felt about the
    learning experience. Your email will be collected throughout these
    forms but will not be used for any purposes and will not be shared
    with anyone but the researcher.\newline Your participation is entirely
    voluntary and if you choose to participate, you can leave the study at
    any time with no penalties. If you choose to leave, your results will
    not be considered. There will be no grades associated with this study,
    therefore it will have no impact on your academic record. The data
    collected from your participation will be recorded, but all personal
    data and identifying information will be kept fully confidential and
    will not be shared with anyone other than the researcher. \\

    \subparagraph{\textbf{Risks and Discomforts}} \hfill

    There is minimal/no risk of discomfort associated with this study.\\

    \subparagraph{\textbf{Benefits}} \hfill

    The information collected from this study may help improve the current
    understanding of how generative AI can be used in the education
    system, and as a result help develop better curricula that prepare
    students better for their futures. \\

    \subparagraph{\textbf{Contact Information}} \hfill

    If you have any questions, you can reach out to me at <researcher
    email>, or my research advisor <advisor email> \\

    \subparagraph{\textbf{Parental Minor Consent (if applicable)}} \hfill

    As a parent or guardian of a minor child, participation in this study
    requires parental consent. You are free to talk to anyone you trust
    about this study and take time to reflect whether you wish your child
    to participate or not. Please read the information about the study
    listed above carefully. \\

    \subparagraph{\textbf{2.}} \textbf{I am a student in an on-level physics or AP Physics 1 class at <name>
        High School or <name> High School, and I (and my parent/legal guardian) have
        read this form and have been able to ask questions of the primary investigator
        and/or discuss my participation with someone I trust. I (and my parent/legal
        guardian) understand that I can ask additional questions at any time during this
        research study and am free to withdraw from participation at any time.}\newline

    \subparagraph{(Mark only one oval)} (*)\hfill \\
    \_\_\_ I am above the age of 18, and I consent to my participation in this survey.\newline
    \_\_\_ I am above the age of 18, and I do not consent to my participation in this survey, and
    I wish to opt out of the survey.\newline
    \_\_\_ I am under the age of a 18, and myself and my parents or guardians have consented
    to my participation in this survey.\newline
    \_\_\_ I am under the age of 18, and myself or my parents or guardian have not consented to my participation in this survey and I wish to opt out of this survey.\newline

    \subparagraph{\textbf{3.}} \textbf{Enter your Participant ID (provided by your teacher). This should follow the format
        XX-0-00 (*)}\\ \vspace{3mm}
    \_\_\_\_\_\_\_\_\_\_\_\_

\subsubsection{GAI-LP Group} \hfill \\
    {\textbf{Email (*): } \_\_\_\_} \\
    \subparagraph{\textbf{INFORMED CONSENT FOR HUMAN PARTICIPATION}} \hfill \\
    {\textbf{Purpose}}

    My name is <researcher> and I am an AP Research Capstone student. In
    this research study, I am interested in looking at how the use of
    generative AI in high school classrooms, particularly with regards to
    STEM subjects (more specifically, on-level physics and AP Physics 1),
    will impact students' educational outcomes. I will be using an
    experimental method, giving students pre- and post-tests along with
    surveys to determine the impact of various learning aides on their
    performance and understanding of a given topic. \\ \hfill

    \subparagraph{\textbf{What can I expect?}} \hfill

    Your participation will involve a short information program regarding
    how to use generative AI responsibly and effectively which will take
    approximately 10 minutes. Then you will take a pre-test on a given
    topic, which should take approximately 10 minutes. After this, the
    study itself will span one full class period (45 minutes). You will
    receive a short lecture (~15 minutes). After this, you will be given
    20 minutes to study this topic using a generative AI, followed by a
    post-test (~10 minutes) to assess any changes in your performance.
    After this, you will also be requested to fill out a short survey
    regarding how you felt about the learning experience. Your email will
    be collected throughout these forms but will not be used for any
    purposes and will not be shared with anyone but the
    researcher.\newline Your participation is entirely voluntary and if
    you choose to participate, you can leave the study at any time with no
    penalties. If you choose to leave, your results will not be
    considered. There will be no grades associated with this study,
    therefore it will have no impact on your academic record. The data
    collected from your participation will be recorded, but all personal
    data and identifying information will be kept fully confidential and
    will not be shared with anyone other than the researcher. \\

    \subparagraph{\textbf{Risks and Discomforts}} \hfill

    There is minimal/no risk of discomfort associated with this study.\\

    \subparagraph{\textbf{Benefits}} \hfill

    The information collected from this study may help improve the current
    understanding of how generative AI can be used in the education
    system, and as a result help develop better curricula that prepare
    students better for their futures. \\

    \subparagraph{\textbf{Contact Information}} \hfill

    If you have any questions, you can reach out to me at <researcher
    email>, or my research advisor <advisor email> \\

    \subparagraph{\textbf{Parental Minor Consent (if applicable)}} \hfill

    As a parent or guardian of a minor child, participation in this study
    requires parental consent. You are free to talk to anyone you trust
    about this study and take time to reflect whether you wish your child
    to participate or not. Please read the information about the study
    listed above carefully. \\

    \subparagraph{\textbf{2.}} \textbf{I am a student in an on-level physics or AP Physics 1 class at <name>
        High School or <name> High School, and I (and my parent/legal guardian) have
        read this form and have been able to ask questions of the primary investigator
        and/or discuss my participation with someone I trust. I (and my parent/legal
        guardian) understand that I can ask additional questions at any time during this
        research study and am free to withdraw from participation at any time.}\newline

    \subparagraph{(Mark only one oval)} (*)\hfill \\
    \_\_\_ I am above the age of 18, and I consent to my participation in this survey.\newline
    \_\_\_ I am above the age of 18, and I do not consent to my participation in this survey, and
    I wish to opt out of the survey.\newline
    \_\_\_ I am under the age of a 18, and myself and my parents or guardians have consented
    to my participation in this survey.\newline
    \_\_\_ I am under the age of 18, and myself or my parents or guardian have not consented to my participation in this survey and I wish to opt out of this survey.\newline

    \subparagraph{\textbf{3.}} \textbf{Enter your Participant ID (provided by your teacher). This should follow the format
        XX-0-00 (*)}\\ \vspace{3mm}
    \_\_\_\_\_\_\_\_\_\_\_\_

    \newpage
    \subsection{Pre/Post-Test Questions}
\subsubsection{Sample Pre/Post-Test Question} \hfill
    \begin{figure}[H]
        \centering
        \includegraphics[width=0.9\linewidth]{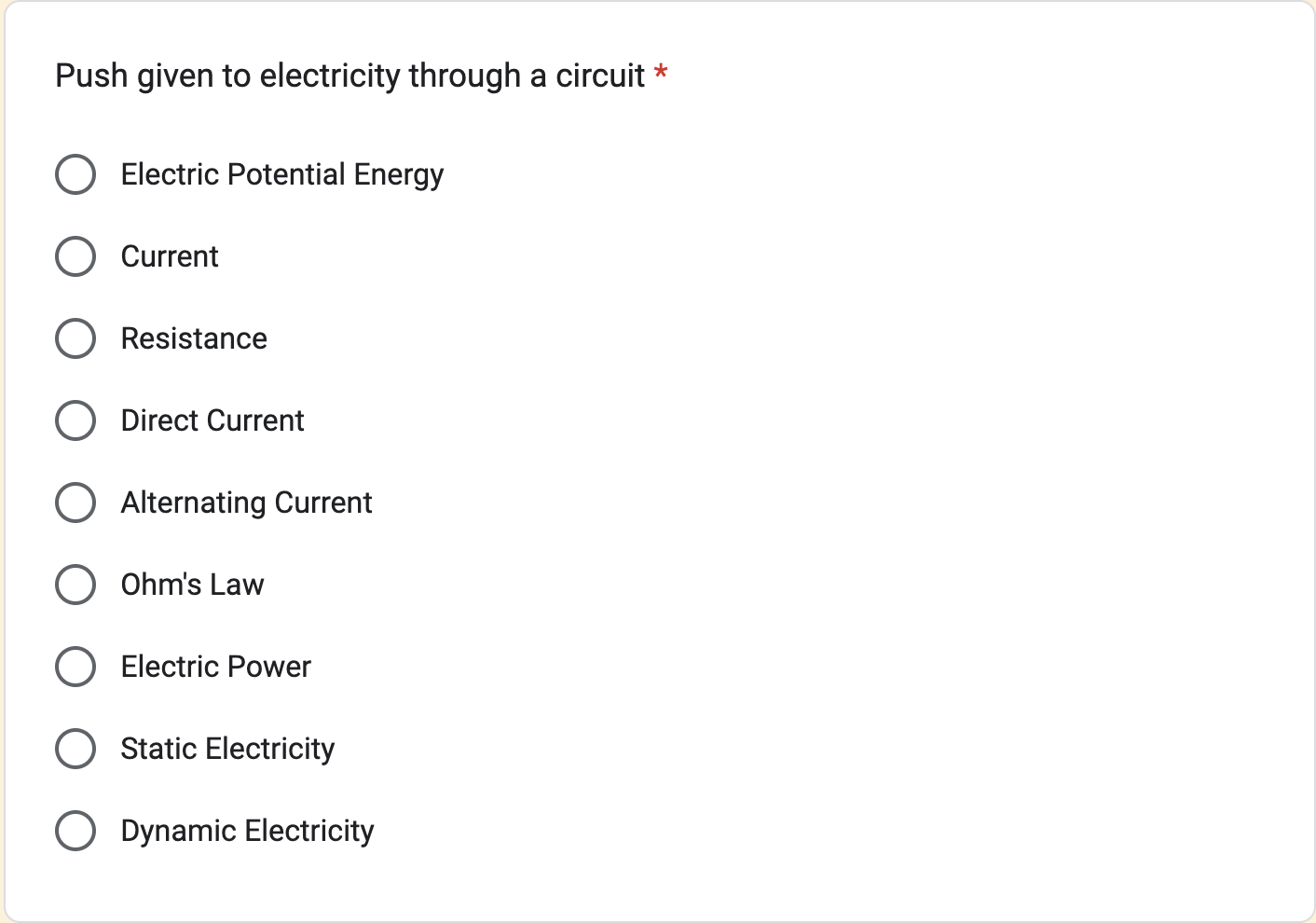}
        \caption{Example Pre/Post-Test Question}
        \label{fig:sample_q_app}
    \end{figure}
\subsubsection{AP Physics Pre-Test} \hfill
    \begin{enumerate}
	    \item Experimental ID:
	    \item Which of the following is NOT a unit of momentum or impulse?
	    \item The direction of an object’s momentum vector is always the
	    	same as the direction of the object’s:
	    \item Why does a momentum square diagram have only quadrants I and
	    	IV?
	    \item What is impulse?
	    \item The impulse due to a varying force is calculated by taking
	    	the:
    \end{enumerate}
\subsubsection{AP Physics Post-Test} \hfill
    \begin{enumerate}
	    \item Experimental ID:
	    \item Car A and Car B have equal masses and travel alongside each
	    	other with equal initial speeds. Car A stops very suddenly, while
	    	Car B gradually comes to rest. Which of the following statements
	    	is NOT true?
	    \item Two balls of equal mass m are dropped from equal heights at
	    	the same time so that each reaches a speed v just before touching
	    	the floor below. Ball A sticks to the floor upon colliding with
	    	it. Ball B bounces back upward to the same height from which it
	    	was dropped. Which of the following gives the correct impulse JA
	    	applied by the floor to ball A and the impulse JB applied by the
	    	floor to ball B?
	    \item A 3 kg object initially moves to the right with a velocity
	    	of +5 m/s. A constant leftward net force of 10 N is applied to
	    	the object for 6 s. At the end of this time, what velocity does
	    	the object have?
	    \item A car speeding up from rest experiences a force that is
	    	initially zero and linearly increases to 1000 N over 4 seconds,
	    	then is a constant value of 1000 N for 2 seconds, then decreases
	    	to zero linearly over 8 seconds. How much momentum did the car
	    	gain?
	    \item A ball of mass m is dropped from rest at the top of a
	    	building and reaches a speed v just before striking the ground.
	    	Which of the following are NOT amounts of impulse that would be
	    	applied to the ball as a result of its collision with the ground,
	    	assuming that no additional energy is added to the ball-Earth
	    	system and that the ground does not collapse? SELECT TWO ANSWERS.
    \end{enumerate}
\subsubsection{On-Level Physics Pre-Test}
    \begin{enumerate}
	    \item Choose the most appropriate term
		    \begin{enumerate}
			    \item Push given to electricity through a circuit:
			    \item Electricity at rest
			    \item Factors that affect resistivity: (select all that
			    	apply)
			    \item Electric current that flows in one direction
			    \item Difference in voltage across a circuit that pushes
			    	electrons through:
		    \end{enumerate}
	    \item While cooking dinner, Karley’s oven uses a 400 V line and
	    	draws 20 A of current when heated to its maximum temperature.
	    	What is the resistance of the oven when it is fully heated?
	    \item An electric heater with a resistance of 40Ω is connected to
	    	a 800-V source. What is the current in the circuit?
    \end{enumerate}
\subsubsection{On-Level Physics Post-Test}
    \begin{enumerate}
	    \item Choose the most appropriate term
		    \begin{enumerate}
			    \item Push given to electricity through a circuit:
			    \item Electricity at rest
			    \item Factors that affect resistivity: (select all that
			    	apply)
			    \item Electric current that flows in one direction
			    \item Difference in voltage across a circuit that pushes
			    	electrons through:
		    \end{enumerate}
	    \item While cooking dinner, Karley’s oven uses a 400 V line and
	    	draws 20 A of current when heated to its maximum temperature.
	    	What is the resistance of the oven when it is fully heated?
	    \item An electric heater with a resistance of 40Ω is connected to
	    	a 800-V source. What is the current in the circuit?
    \end{enumerate}
    \begin{description}
        \item[\textit{Note:}] The On-Level Physics Pre- and Post-Tests were identical. However, students were not able to see the correct answers or which questions they missed.
    \end{description}
    \newpage
    \subsection{Survey (Post)}
    

\section*{Supplementary material}

# Generative AI in STEM Education

* Indicates required question

1. Email *

   _____

2. Please enter your experimental ID: *

   _____

3. How would you rate your understanding of this topic? *

   *Mark only one oval.*

   |     | 1 | 2 | 3 | 4 | 5 | 6 | 7 | 8 | 9 | 10 |                    |
   |-----|---|---|---|---|---|---|---|---|---|----|--------------------|
   | Not | ○ | ○ | ○ | ○ | ○ | ○ | ○ | ○ | ○ | ○  | Extremely Confident |

4. How do you feel you performed on the post-assessment? *

   *Mark only one oval.*

   |     | 1 | 2 | 3 | 4 | 5 | 6 | 7 | 8 | 9 | 10 |           |
   |-----|---|---|---|---|---|---|---|---|---|----|-----------|
   | Poo | ○ | ○ | ○ | ○ | ○ | ○ | ○ | ○ | ○ | ○  | Excellent |

5. How much did you enjoy learning this topic? *

   *Mark only one oval.*

   |   | 1 | 2 | 3 | 4 | 5 |   |
   |---|---|---|---|---|---|---|
   | Stro | ○ | ○ | ○ | ○ | ○ | Strong Like |

6. How engaged did you feel during the learning process? *

   *Mark only one oval.*

   |   | 1 | 2 | 3 | 4 | 5 |   |
   |---|---|---|---|---|---|---|
   | Not | ○ | ○ | ○ | ○ | ○ | Very Engaged |

7. Class Period *

   *Mark only one oval.*

   | | |
   |---|---|
   | ○ TMHS - AP Physics 1 (Hunnicutt) - 4th Period | *Skip to question 25* |
   | ○ TMHS - AP Physics 1 (Hunnicutt) - 5th Period | *Skip to question 8* |
   | ○ TMHS - AP Physics 1 (Hunnicutt) - 6th period | *Skip to question 14* |
   | ○ TMHS - On-Level Physics (Kilgore) - 4th Period | *Skip to question 25* |
   | ○ TMHS - On-Level Physics (Kilgore) - 5th Period | *Skip to question 8* |
   | ○ TMHS - On-Level Physics (Kilgore) - 6th Period | *Skip to question 14* |
   | ○ TMHS - On-Level Physics (Kilgore) - 7th Period | *Skip to question 14* |

Search Engine

Exploring the Applications of Generative AI in High School STEM Education    278. How would you rate the effectiveness of this tool in improving your understanding *
   of this topic?

   *Mark only one oval.*

   |   | 1 | 2 | 3 | 4 | 5 |   |
   |---|---|---|---|---|---|---|
   | Not | ◯ | ◯ | ◯ | ◯ | ◯ | Very helpful |

9. Did you feel that this tool impacted your learning? *

   *Mark only one oval.*

   |   | 1 | 2 | 3 | 4 | 5 |   |
   |---|---|---|---|---|---|---|
   | Stro | ◯ | ◯ | ◯ | ◯ | ◯ | Strong Positive Impact |

10. Did you feel that using this tool improved your confidence with this topic? *

    *Mark only one oval.*

    |   | 1 | 2 | 3 | 4 | 5 |   |
    |---|---|---|---|---|---|---|
    | Stro | ◯ | ◯ | ◯ | ◯ | ◯ | Strong Positive Impact |

11. Did you feel that using this tool impacted your enjoyment of learning this topic? *

    *Mark only one oval.*

    |   | 1 | 2 | 3 | 4 | 5 |   |
    |---|---|---|---|---|---|---|
    | Stro | ◯ | ◯ | ◯ | ◯ | ◯ | Strong Positive Impact |

Manuscript submitted to ACM

12. Did you feel that the information you gathered from the search engine was helpful? *

    *Mark only one oval.*

    |   | 1 | 2 | 3 | 4 | 5 |   |
    |---|---|---|---|---|---|---|
    | Not | ○ | ○ | ○ | ○ | ○ | Very Helpful |

13. How would you rate the factual accuracy of the results that you found using the search engine? *

    *Mark only one oval.*

    |   | 1 | 2 | 3 | 4 | 5 |   |
    |---|---|---|---|---|---|---|
    | High | ○ | ○ | ○ | ○ | ○ | Highly Accurate |

    *Skip to question 25*

    Generative AI

14. How would you rate the effectiveness of this tool in improving your understanding of this topic? *

    *Mark only one oval.*

    |   | 1 | 2 | 3 | 4 | 5 |   |
    |---|---|---|---|---|---|---|
    | Not | ○ | ○ | ○ | ○ | ○ | Very helpful |

15. Did you feel that this tool impacted your learning? *

    *Mark only one oval.*

    |  | 1 | 2 | 3 | 4 | 5 |  |
    |---|---|---|---|---|---|---|
    | Stro | ○ | ○ | ○ | ○ | ○ | Strong Positive Impact |

16. Did you feel that using this tool improved your confidence with this topic? *

    *Mark only one oval.*

    |  | 1 | 2 | 3 | 4 | 5 |  |
    |---|---|---|---|---|---|---|
    | Stro | ○ | ○ | ○ | ○ | ○ | Strong Positive Impact |

17. Did you feel that using this tool impacted your enjoyment of learning this topic? *

    *Mark only one oval.*

    |  | 1 | 2 | 3 | 4 | 5 |  |
    |---|---|---|---|---|---|---|
    | Stro | ○ | ○ | ○ | ○ | ○ | Strong Positive Impact |

18. Did you feel that the information outputted by the generative AI was helpful? *

    *Mark only one oval.*

    |  | 1 | 2 | 3 | 4 | 5 |  |
    |---|---|---|---|---|---|---|
    | Not | ○ | ○ | ○ | ○ | ○ | Very Helpful |

19. How would you rate the factual accuracy of the information outputted by the generative AI? *

    *Mark only one oval.*

    1  2  3  4  5

    High ⃝ ⃝ ⃝ ⃝ ⃝ Highly Accurate

20. (Optional) Do you have any other comments regarding the generative AI?

    _____
    _____
    _____
    _____

21. Were you a part of the generative AI literacy program? *

    *Mark only one oval.*

    ⃝ Yes (Kilgore 7th period; Hunnicutt 6th period - slideshow)

    ⃝ No    *Skip to question 25*

    *Skip to question 25*

    Generative AI Literacy Program

22. How did you feel the literacy program impacted your confidence in your ability to use the generative AI? *

    *Mark only one oval.*

    1  2  3  4  5

    Stro ⃝ ⃝ ⃝ ⃝ ⃝ Strong Positive Impact

23. Did you feel that the literacy program improved your understanding of generative AI? *

    *Mark only one oval.*

    | | 1 | 2 | 3 | 4 | 5 | |
    |---|---|---|---|---|---|---|
    | No I | ○ | ○ | ○ | ○ | ○ | Strong Improvement |

24. (Optional) Do you have any additional concerns or feedback regarding the generative AI literacy program?

    _____
    _____
    _____
    _____
    _____

    *Skip to question 25*

    End

25. (Optional) Did you have any other remarks/feedback?

    _____
    _____
    _____
    _____
    _____

This content is neither created nor endorsed by Google.

Google Forms



    \section{Data Results}
    \subsection{AP Physics Test Score Differences}
\subsubsection{Descriptives Table} \hfill
    \begin{table}[h]
        \caption{Descriptive Statistics}
        \label{tab:ap_fulldescriptive}
        {
            \begin{tabular}{lrrrr}
                \toprule
                \multicolumn{1}{c}{} & \multicolumn{4}{c}{Difference}                                  \\
                \cline{2-5}
                                     & Control                        & SE       & GAI      & GAI-LP   \\
                \cmidrule[0.4pt]{1-5}
                Valid                & $26$                           & $31$     & $14$     & $14$     \\
                Mean                 & $-0.346$                       & $0.484$  & $-0.857$ & $0.500$  \\
                95\% CI Mean Upper   & $0.201$                        & $0.993$  & $0.177$  & $1.207$  \\
                95\% CI Mean Lower   & $-0.893$                       & $-0.025$ & $-1.891$ & $-0.207$ \\
                Std. Deviation       & $1.355$                        & $1.387$  & $1.791$  & $1.225$  \\
                Variance             & $1.835$                        & $1.925$  & $3.209$  & $1.500$  \\
                Minimum              & $-3.000$                       & $-3.000$ & $-4.000$ & $-2.000$ \\
                Maximum              & $3.000$                        & $4.000$  & $2.000$  & $3.000$  \\
                \bottomrule
            \end{tabular}
        }
    \end{table}

\subsubsection{Distribution Plots} \hfill \\
    \begin{figure}[H]
        \centering
        \begin{subfigure}{0.45\textwidth}
            \includegraphics[width=0.9\linewidth]{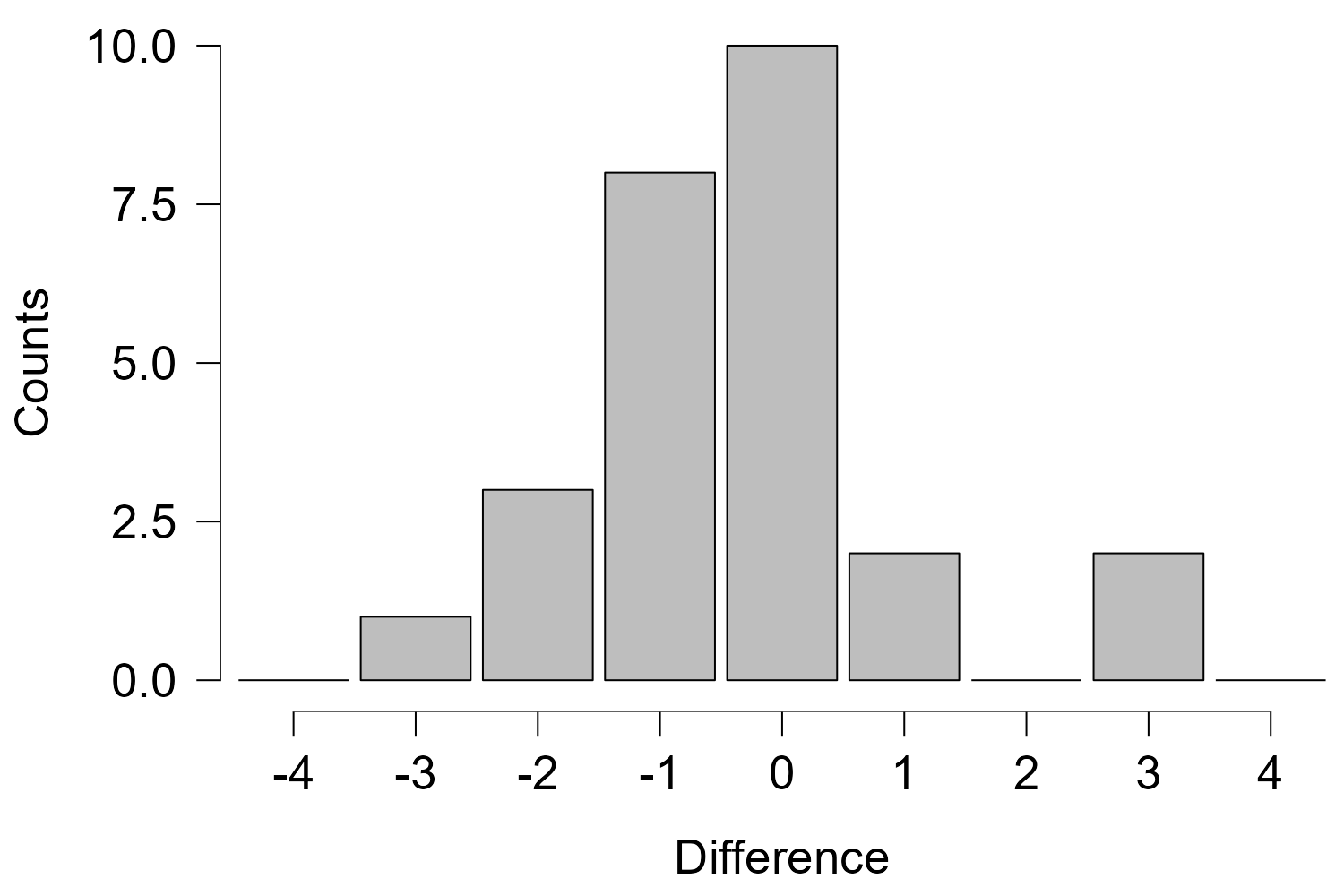}
            \caption{Control}
            \label{fig:ap-control-app}
        \end{subfigure}
        \begin{subfigure}{0.45\textwidth}
            \includegraphics[width=0.9\linewidth]{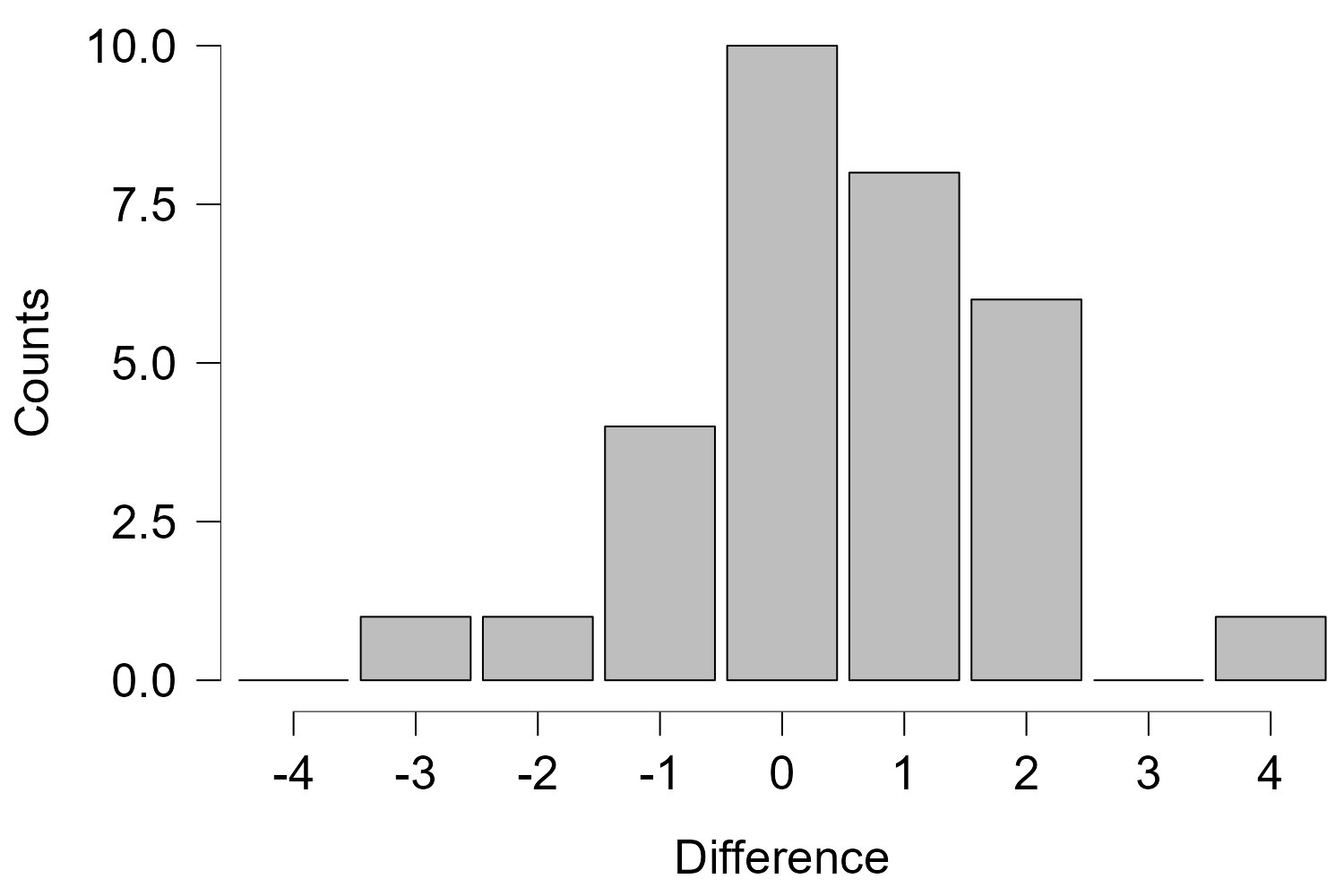}
            \caption{SE}
            \label{fig:ap-se-app}
        \end{subfigure}
        \begin{subfigure}{0.45\textwidth}
            \includegraphics[width=0.9\linewidth]{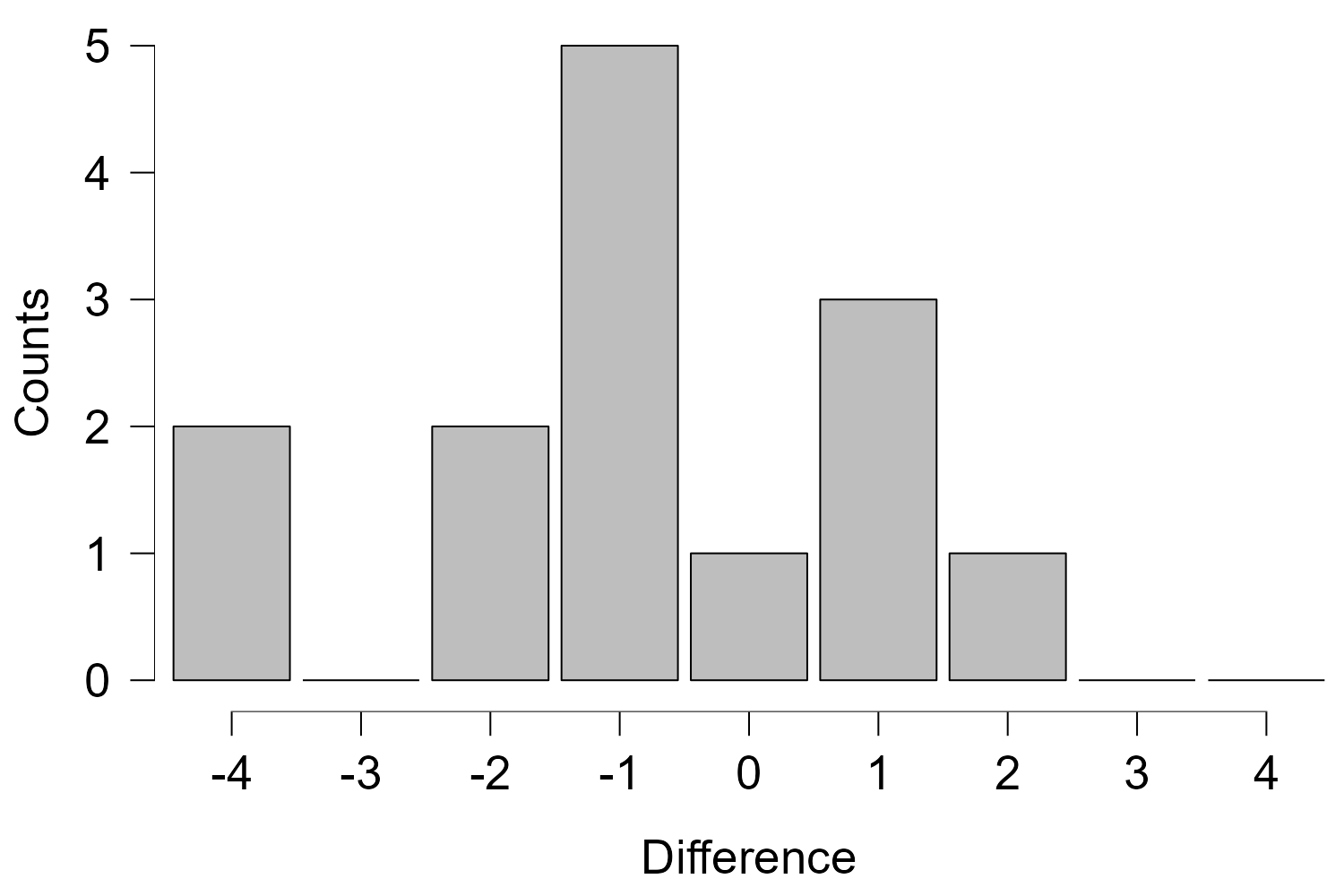}
            \caption{GAI}
            \label{fig:ap-gai-app}
        \end{subfigure}
        \begin{subfigure}{0.45\textwidth}
            \includegraphics[width=0.9\linewidth]{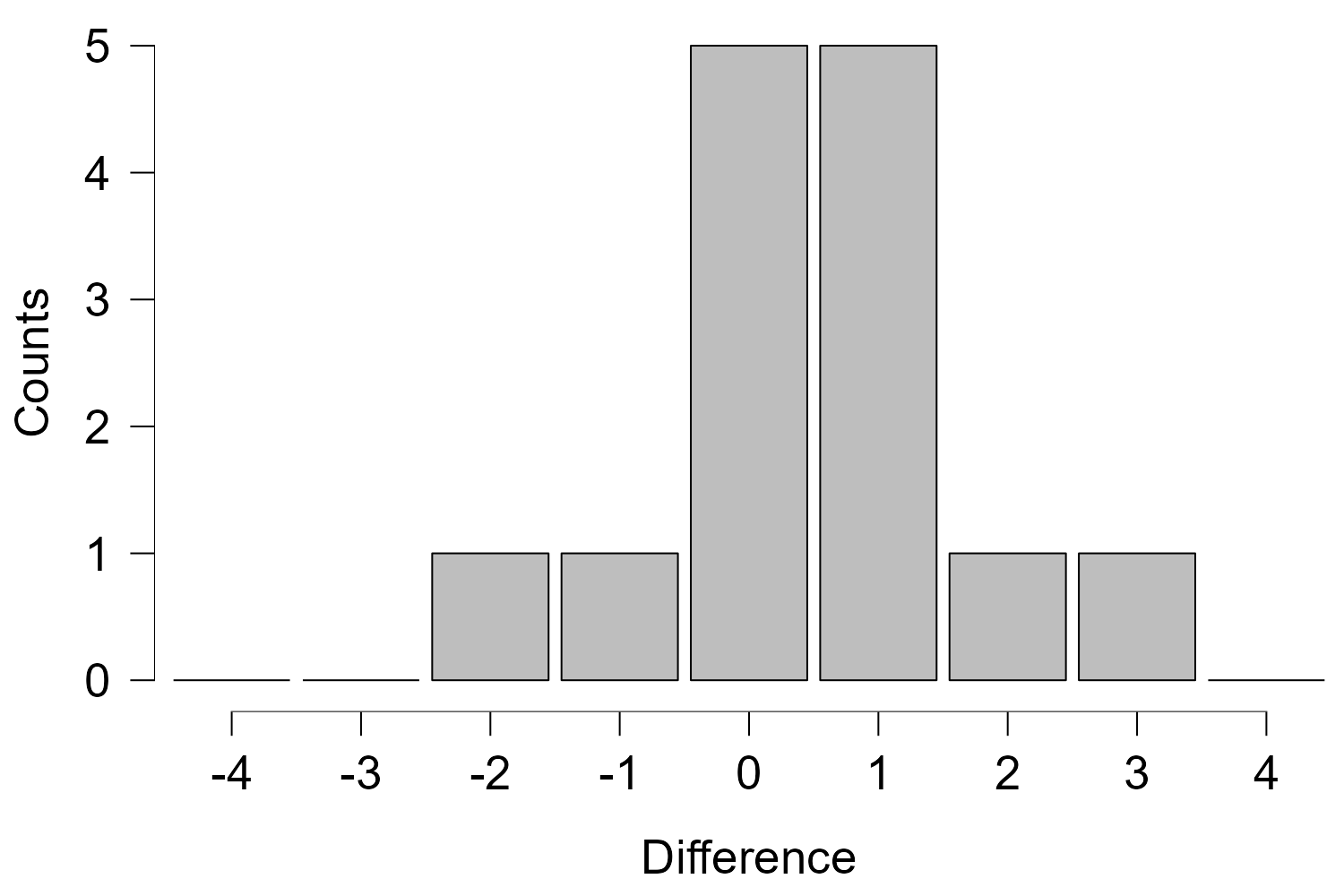}
            \caption{GAI-LP}
            \label{fig:ap-gailp-app}
        \end{subfigure}
        \caption{Distribution Plots for AP Physics Test Score Differences}
        \label{fig:ap-plots}
    \end{figure}

\subsubsection{ANOVA Test Results} \hfill \\
    \begin{table}[H]
        \centering
        \caption{ANOVA - Difference}
        \label{tab:ap-anova-Difference}
        {
            \begin{tabular}{lrrrrrrrr}
                \toprule
                Cases              & Sum of Squares & df   & Mean Square & F       & p       & $\eta$$^{2}$ & $\eta$$^{2}_p$ & $\omega$$^{2}$ \\
                    \cmidrule[0.4pt]{1-9}
                Experimental Group & $24.147$       & $3$  & $8.049$     & $3.955$ & $0.011$ & $0.128$      & $0.128$        & $0.094$        \\
                Residuals          & $164.841$      & $81$ & $2.035$     & $$      & $ $     & $$           & $$             & $$             \\
                \bottomrule
            \end{tabular}
        }
    \end{table}

    \vspace{5mm}
    \begin{table}[h]
        \centering
        \caption{Simple Contrast - Experimental Group}
        \label{tab:simpleContrast-ExperimentalGroup}
        {
            \begin{tabular}{lrrrrrrr}
                \toprule
                \multicolumn{1}{c}{} & \multicolumn{1}{c}{} & \multicolumn{2}{c}{95\% CI for Mean Difference} & \multicolumn{1}{c}{} & \multicolumn{1}{c}{} & \multicolumn{1}{c}{} & \multicolumn{1}{c}{}           \\
                \cline{3-4}
                Comparison           & Estimate             & Lower                                           & Upper                & SE                   & df                   & t                    & p       \\
                \cmidrule[0.4pt]{1-8}
                SE - Control         & $0.830$              & $0.075$                                         & $1.585$              & $0.379$              & $81$                 & $2.188$              & $0.032$ \\
                GAI - Control        & $-0.511$             & $-1.452$                                        & $0.430$              & $0.473$              & $81$                 & $-1.081$             & $0.283$ \\
                GAI-LP - Control     & $0.846$              & $-0.095$                                        & $1.787$              & $0.473$              & $81$                 & $1.789$              & $0.077$ \\
                \bottomrule
            \end{tabular}
        }
    \end{table}

    \subsection{On-Level Physics Test Score Differences}
\subsubsection{Descriptives Table} \hfill
    \begin{table}[h]
        \centering
        \caption{Descriptive Statistics}
        \label{tab:descriptiveStatistics}
        {
            \begin{tabular}{lrrrr}
                \toprule
                \multicolumn{1}{c}{} & \multicolumn{4}{c}{Difference}                                  \\
                \cline{2-5}
                                     & Control                        & SE       & GAI      & GAI-LP   \\
                \cmidrule[0.4pt]{1-5}
                Valid                & $21$                           & $16$     & $14$     & $21$     \\
                Mean                 & $0.905$                        & $1.563$  & $0.714$  & $1.190$  \\
                95\% CI Mean Upper   & $1.679$                        & $2.761$  & $1.984$  & $1.905$  \\
                95\% CI Mean Lower   & $0.131$                        & $0.364$  & $-0.555$ & $0.476$  \\
                Std. Deviation       & $1.700$                        & $2.250$  & $2.199$  & $1.569$  \\
                Variance             & $2.890$                        & $5.063$  & $4.835$  & $2.462$  \\
                Minimum              & $-2.000$                       & $-1.000$ & $-4.000$ & $-2.000$ \\
                Maximum              & $3.000$                        & $6.000$  & $4.000$  & $4.000$  \\
                \bottomrule
            \end{tabular}
        }
    \end{table}

\subsubsection{Distribution Plots} \hfill \\
    \begin{figure}[H]
        \centering
        \begin{subfigure}{0.45\textwidth}
            \includegraphics[width=0.9\linewidth]{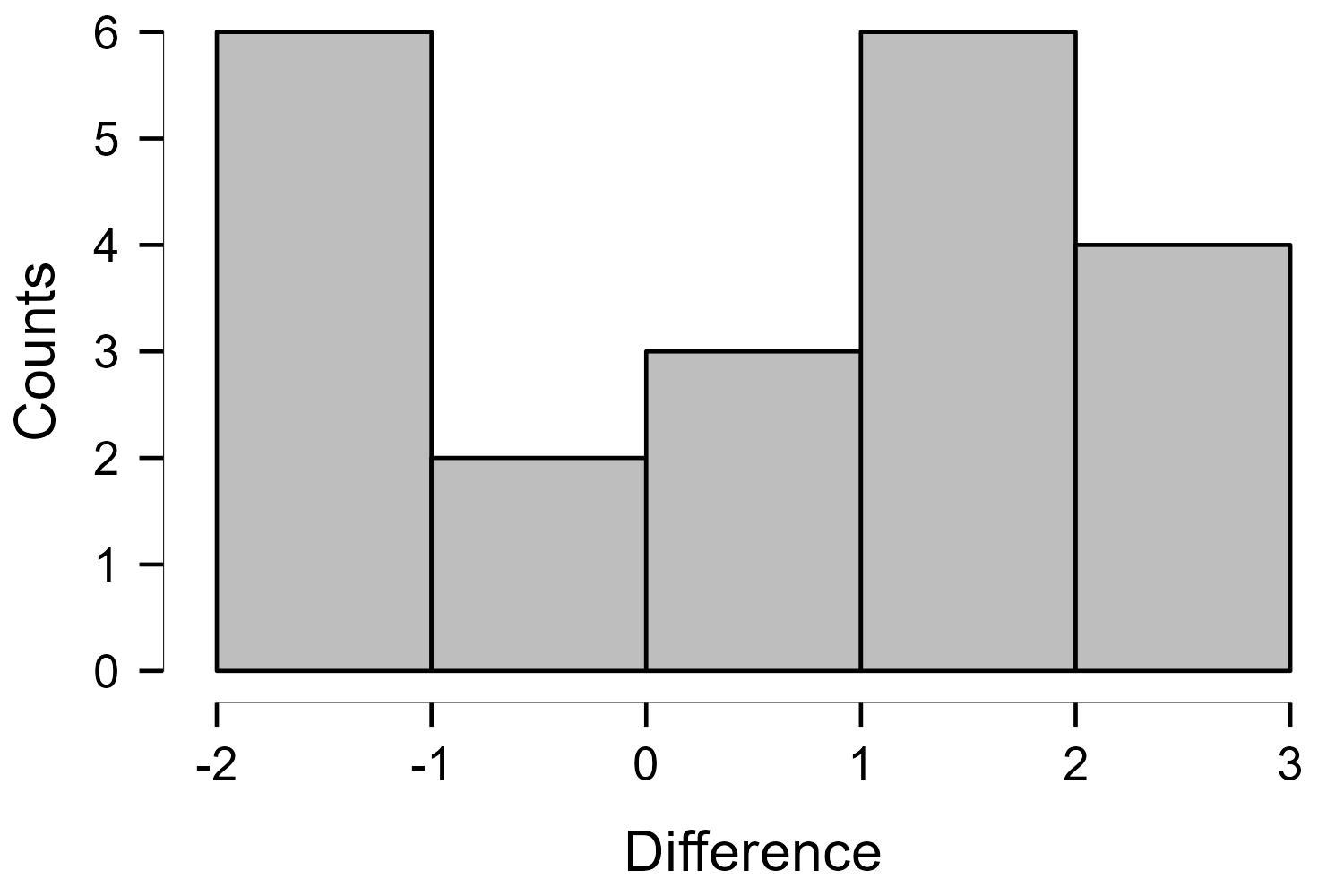}
            \caption{Control}
            \label{fig:ol-control-app}
        \end{subfigure}
        \begin{subfigure}{0.45\textwidth}
            \includegraphics[width=0.9\linewidth]{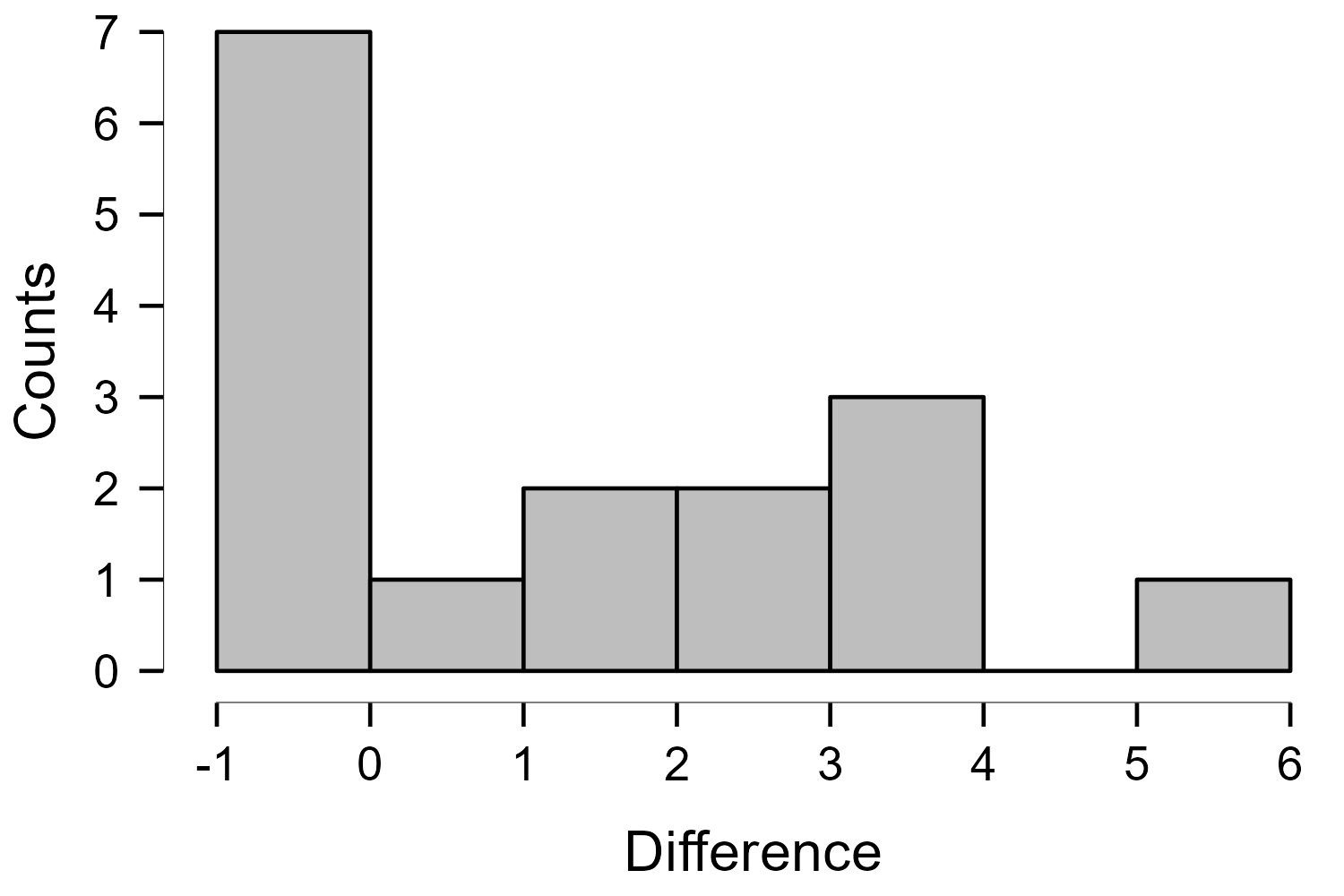}
            \caption{SE}
            \label{fig:ol-se-app}
        \end{subfigure}
        \begin{subfigure}{0.45\textwidth}
            \includegraphics[width=0.9\linewidth]{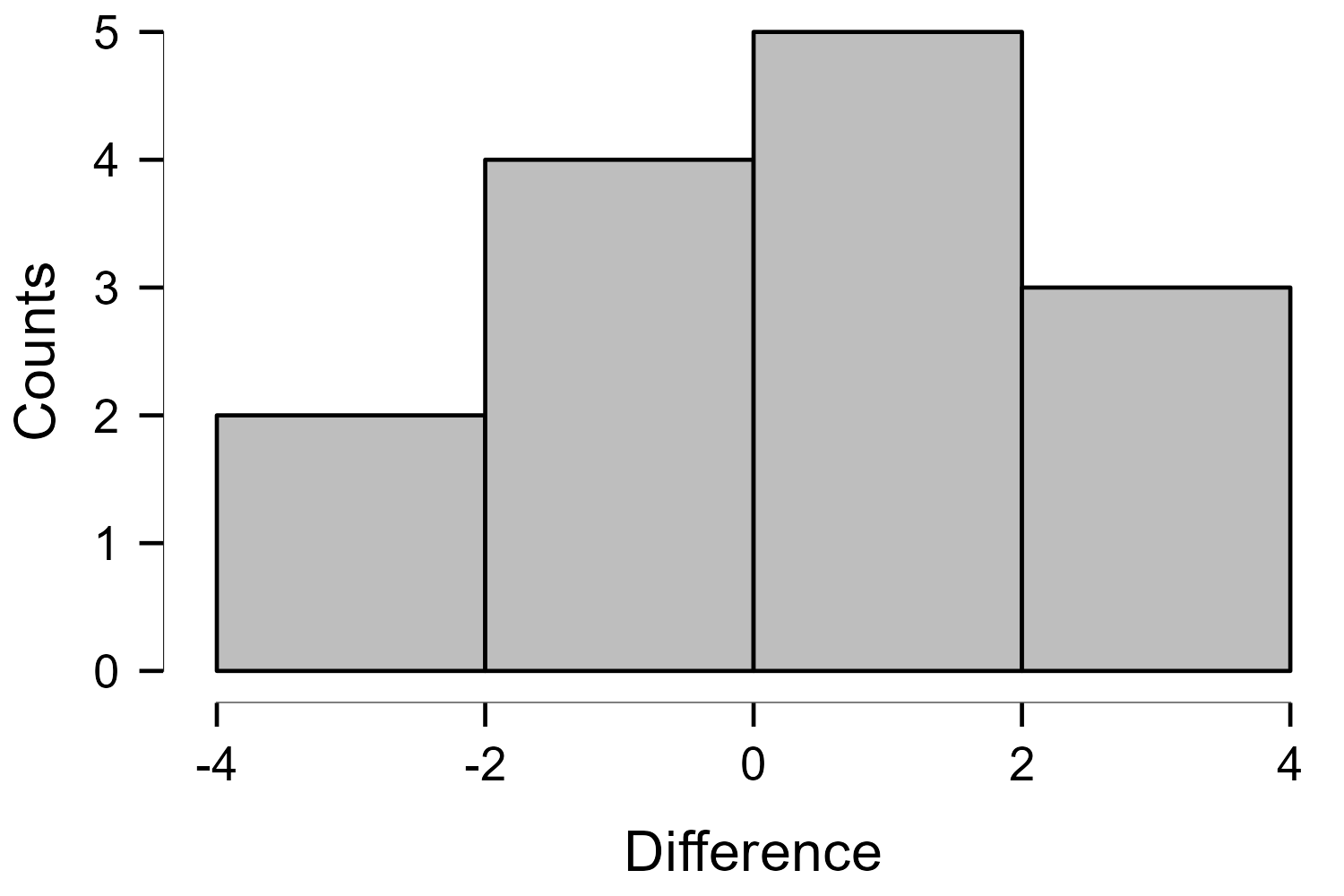}
            \caption{GAI}
            \label{fig:ol-gai-app}
        \end{subfigure}
        \begin{subfigure}{0.45\textwidth}
            \includegraphics[width=0.9\linewidth]{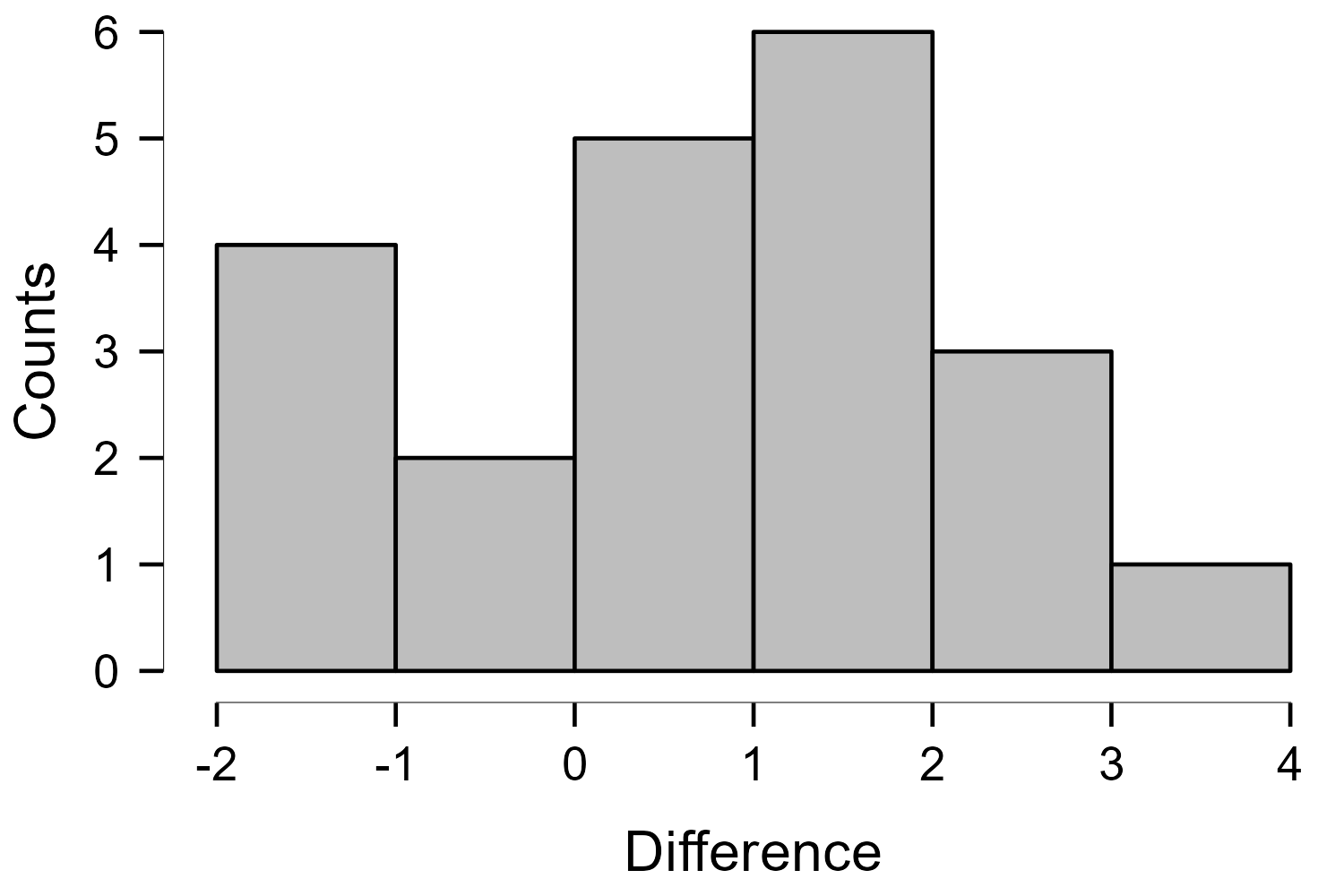}
            \caption{GAI-LP}
            \label{fig:ol-gailp-app}
        \end{subfigure}
        \caption{Distribution Plots for On-Level Physics Test Score Differences}
        \label{fig:ol-plots}
    \end{figure}
    \newpage
\subsubsection{ANOVA Test Results} \hfill \\
    \begin{table}[h]
        \centering
        \caption{ANOVA - Difference}
        \label{tab:ol-anova-difference}
        {
            \begin{tabular}{lrrrrrrrr}
                \toprule
                Cases     & Sum of Squares & df   & Mean Square & F       & p       & $\eta$$^{2}$ & $\eta$$^{2}_p$ & $\omega$$^{2}$ \\
                    \cmidrule[0.4pt]{1-9}
                Treatment & $6.477$        & $3$  & $2.159$     & $0.597$ & $0.619$ & $0.026$      & $0.026$        & $0.000$        \\
                Residuals & $245.842$      & $68$ & $3.615$     & $$      & $ $     & $$           & $$             & $$             \\
                \bottomrule
            \end{tabular}
        }
    \end{table}

    \vspace{5mm}
    \begin{table}[h]
        \centering
        \caption{Simple Contrast - Treatment}
        \label{tab:simpleContrast-Treatment}
        {
            \begin{tabular}{lrrrrr}
                \toprule
                Comparison       & Estimate & SE      & df   & t        & p       \\
                \cmidrule[0.4pt]{1-6}
                SE - Control     & $0.658$  & $0.631$ & $68$ & $1.042$  & $0.301$ \\
                GAI - Control    & $-0.190$ & $0.656$ & $68$ & $-0.290$ & $0.772$ \\
                GAI-LP - Control & $0.286$  & $0.587$ & $68$ & $0.487$  & $0.628$ \\
                \bottomrule
            \end{tabular}
        }
    \end{table}

    \newpage
    \section{Author Statement}

    This submission to TOCHI has no relation to any previous papers of the
    author.

\end{document}